\documentclass[fleqn,usenatbib]{mnras}

\usepackage[T1]{fontenc}

\DeclareRobustCommand{\VAN}[3]{#2}
\let\VANthebibliography\thebibliography
\def\thebibliography{\DeclareRobustCommand{\VAN}[3]{##3}\VANthebibliography}

\usepackage{graphicx}	% Including figure files
\usepackage{amsmath}	% Advanced maths commands
\usepackage{amssymb}	% Extra maths symbols

\usepackage{xcolor}
\usepackage[normalem]{ulem}
\newcommand{\tpm}{t$_\text{PM}$}

\newcommand{\QPM}{Q$_\text{PM}$}
\newcommand{\QCTRL}{Q$_\text{CTRL}$}
\newcommand{\SFPM}{SF$_\text{PM}$}
\newcommand{\SFCTRL}{SF$_\text{CTRL}$}

\newcommand{\rgal}{R$_\text{gal}$}
\newcommand{\msol}{M$_\odot$}
\newcommand{\bhm}{M$_\text{BH}$}

\usepackage{newtxtext,newtxmath}

\title[Quenching in TNG post-mergers.]{Interacting galaxies in the IllustrisTNG simulations - III: (the rarity of) quenching in post-merger galaxies.}

\author[S. Quai et al.]{
Salvatore Quai,$^{1}$\thanks{E-mail: squai@uvic.ca}, Maan H. Hani,$^{1,2}$\thanks{Herschel Fellow} Sara L. Ellison,$^{1}$ David R. Patton,$^{3}$ Joanna Woo$^{4}$
\\
$^{1}$Department of Physics and Astronomy, University of Victoria, 3800 Finnerty Rd, Victoria, BC V8P 5C2, Canada\\
$^{2}$ Department of Physics and Astronomy, McMaster University Hamilton ON L8S 4M1, Canada\\
$^{3}$ Department of Physics and Astronomy, Trent University, 1600 West Bank Drive, Peterborough, ON K9L 0G2, Canada\\
$^{4}$ Department of Physics, Simon Fraser University, 8888 University Dr, Burnby BC, V5A 1S6, Canada\\
}

\date{Accepted XXX. Received YYY; in original form ZZZ}

\pubyear{2020}

\begin{document}
\label{firstpage}
\pagerange{\pageref{firstpage}--\pageref{lastpage}}
\maketitle

\begin{abstract}
Galaxy mergers are traditionally one of the favoured mechanisms for the transformation of spiral galaxies to spheroids and for quenching star formation. 
To test this paradigm in the context of modern cosmological simulations, we use the IllustrisTNG simulation to investigate the impact of individual merger events on quenching star formation (i.e. star formation rate at least $3\sigma$ below the star-forming main sequence) within $500$ Myr after the coalescence phase. 
The rate of quenching
 amongst recently merged galaxies is compared with a control sample that is matched in redshift, stellar mass, star formation rate (SFR), black hole mass and environment. 
We find quenching to be uncommon among the descendants of post-merger galaxies, with only $\sim5\%$ of galaxies quenching within $500$ Myr after the merger. 
Despite this low absolute rate, we find that quenching occurs in post-mergers at twice the rate of the control galaxies. 
The fraction of quenched post-merger descendants $1.5$ Gyr after the merger becomes statistically indistinguishable from that of non-post-mergers,  suggesting that mergers could speed up the quenching process in those post-mergers whose progenitors had physical conditions able to sustain effective active galactic nuclei (AGN) kinetic feedback, thus capable of removing gas from galaxies.
Our results indicate that although quenching does not commonly occur promptly after coalescence, mergers nonetheless do promote the cessation of star formation in some post-mergers.
We find that, in IllustrisTNG, it is the implementation of the AGN kinetic feedback that is responsible for quenching post-mergers, as well as non-post-merger controls. %As soon as kinetic energy balances the gravitational binding energy of the gas, galaxies start experiencing gas loss, and eventually, their star formation quenches when the gas fraction drops below $\sim0.1$.
As a result of the released kinetic energy, galaxies experience gas loss and eventually, they will quench. 
Galaxies with an initially low gas fraction show a preferable pre-disposition towards quenching.  
The primary distinguishing factor between quenched and star-forming galaxies is gas fraction, with a sharp boundary at f$_\text{gas}\sim0.1$ in TNG.
%There is a pre-disposition that favours quenching in those galaxies whose gas fraction is low, and the gas fraction results to be a very good discriminator between the quenched and star-forming populations: we find a very sharp distinction in gas fraction, with the quenched population dominating at gas fractions below f$_\text{gas} \sim 0.1$.}
\end{abstract}


%%%%%%%%%%%%%%%%%%%%%%%%%%%%%%%%%%%%%%%%%%%%%%%%%%

%%%%%%%%%%%%%%%%% BODY OF PAPER %%%%%%%%%%%%%%%%%%

\section{Introduction}
Galaxies show bimodal distributions in fundamental properties, such as colours and structure, both locally \citep[e.g.,][]{Strateva2001, Blanton2003,  Baldry2004, Kauffmann2003} and out to at least $\text{z} \sim2$, \citep[e.g.,][]{Bell2004, Willmer2006, Whitaker2011, Wuyts2011, Bell2012, Cheung2012, Muzzin2013}. 
Moreover, there is strong evidence that there is continuous growth in the number density and stellar mass of the red and passively evolving early-type population from z$\sim1 - 2$ to the present \citep[e.g.,][]{Bell2004, Faber2007, Pozzetti2010, Ilbert2013}. 
This implies that a large fraction of late-type galaxies convert into early-types, due to the suppression of star formation (hereafter quenching) that accompanies their change in morphologies \citep[e.g.,][]{Drory2004, Faber2007,  Peng2010, Pozzetti2010, Tacchella2015, Woo2015}.
It is also thought that these transitional scenarios depend on the environment in which the galaxies are located \citep[e.g.,][]{Goto2003, Balogh2004, Peng2010, Woo2017}. 

Many mechanisms have been proposed for the physical origin of star formation quenching \citep[see for example the review by][]{Somerville2015}, including (i) the heating of the inner halo gas by cosmological accretion via ram pressure drag and local shocks \citep[i.e. gravitational quenching,][]{Dekel2008}, (ii) the stability of discs against fragmentation to bound clumps \citep[i.e. morphological quenching,][]{Martig2009, Gensior2020a, Gensior2020b}, (iii) the removal of the gas supply due to active galactic nuclei (AGN) activity, and/or stellar feedback \citep[e.g.,][]{DiMatteo2005, Bower2006, Sijacki2007, Cattaneo2009, Fabian2012}, (iv) the interaction between the galaxy gas with the intracluster medium in high density environments \citep[i.e. environmental or satellite quenching,][]{Gunn1972, Larson1980, Moore1998, Bekki2009, Peng2010, Peng2012}, and (v) the interaction with other galaxies \citep[i.e. major mergers,][]{DiMatteo2005,  Springel2005b, Springel2005a, Croton2006, Hopkins2008, Somerville2008}.
Quenching processes tend to be classified as internal or environmental depending on whether they originated within a galaxy or if they are triggered by the influence of the external factors (e.g. the intra-cluster medium). These processes are not mutually exclusive, and they could in principle take place together on different timescales. 
%For instance, the environmental quenching (e.g. gas stripping) is expected to be dominant only in dense groups and clusters. 

AGN quenching is one of the most popularly invoked mechanisms for quenching star formation, but it remains controversial.
On the one hand, several observational studies have supported the thesis that AGN feedback should be able to remove gas from the galactic reservoir, eventually leading to quenching \citep[e.g.,][]{ Kaviraj2007, Fabian2012, Cimatti2013}. 
The link between AGN and quenching is also supported by the theoretical results obtained combining N-body simulations of dark matter halo evolution \citep{Springel2005, BoylanKolchin2009} with semi-analytic models for galaxy formation \citep{White1991, Springel2005, Lu2011, Benson2012}. 
On the other hand, there is a growing body of literature that finds that AGN have normal gas reservoirs, both in the atomic \citep[e.g.,][]{Ellison2019}, and molecular gas-phase \citep[e.g.,][]{Shangguan2018, Shangguan2020, Koss2020, Jarvis2020}, with the possible exception of gas depletion in the dwarf regime \citep[][]{Bradford2018, Ellison2019}.
The persistence of large gas reservoirs is at odds with the scenario of AGN driven feedback leading to quenching.

Since many models of galaxy interactions lead to AGN triggering, it has long been suggested that galaxy mergers could provide a major pathway for galaxy quenching \citep{DiMatteo2005, Springel2005b, Springel2005a, Croton2006, Hopkins2008, Somerville2008}. 
%From a theoretical point of view, many simulations of mergers predict strong inflows of gas occurring both during close encounters between galaxy pairs and during the coalescence phase \citep[][, other citations]{Hernquist1989}.
%This can lead to enhanced star formation, with starburst episodes, diluted gas-phase metallicities and increased AGN activity due to accretions onto the central black hole \citep[e.g.,][other citations]{Hernquist1989, Mihos1996, Torrey2012, Hani2019}. 
Strong mechanical AGN feedback triggered by the inflow of gas at low angular momentum, could potentially drive out the gas from the galaxy. 
This halts the star formation, and hampers the replenishment of the galactic gas reservoir. 
At the same time, theory predicts strong morphological disturbances \citep[e.g.,][]{DiMatteo2007} that should accompany the migration from the blue cloud to the red sequence \citep[][]{Mihos1996, Hopkins2008, Somerville2008} consistent with observations that quenched galaxies tend to have spheroidal morphologies \citep[e.g., the review by][]{Conselice2014}.
%On the one hand, several observational studies of merging galaxies have supported the theoretical predictions of structural asymmetries, enhanced star formation, central dilution of gas-phase metallicity and intensified AGN activities \citep[][other citations]{Patton2005, Ellison2008, Ellison2010, Scudder2012, Patton2013, Ellison2013, Ellison2019, Thorp2019, Patton2020}.
%On the other hand, the relation between mergers and quenching remains uncertain.
%For example, by analysing a sample extracted from the SDSS DR7, \citet{Weigel2017} argued that major mergers should not be the preferred path leading to permanent quenching. They found that major merger quenched galaxies account for a maximum of 5\% of the quenched population at a given stellar mass, both at low- and intermediate-redshift. 

Given the advances in simulations over the last decade, it is worth reviewing the predicted link between mergers, AGN triggering and quenching.
Previous simulations that linked mergers to AGN activity tended to have very aggressive feedback recipes \citep[e.g.,][]{Springel2005a, Bower2006, Khalatyan2008, Hopkins2008}.
However, we currently know from observations that the majority of interacting galaxies (i.e.mergers and pairs) in the local universe show a relatively modest enhancement in both star formation and AGN luminosity \citep[][]{Patton2005, Li2008, Ellison2008, Jogee2009, Scudder2012, Patton2013, Ellison2013, Rodighiero2015, Knapen2015, Ellison2019, Thorp2019, Patton2020}.
Both theoretical and observational studies agree on an increase of the major merger rate with redshift out to at least z $\sim1.5$ \citep[e.g.][]{Lin2008, DeRavel2009, Lotz2011, LopezSanjuan2013, RodriguezGomez2015}, whilst minor mergers show little evolution with redshift \citep[e.g.][]{Lotz2011}. However, there is no consensus on whether the contribution of galaxy mergers to star formation decreases with increasing redshift \citep[e.g.][]{Rodighiero2011, Wilson2019}, or continue to produce enhanced star formation \citep[e.g.][]{Lin2007, Wong2011}.
Moreover, there has been no previous work to study the quenching statistics of mergers in a full cosmological setting. 
The work presented here investigates merger driven quenching in an unbiased statistical galaxy sample, using a modern AGN implementation offered by the IllustrisTNG simulations \citep{Springel2018, Nelson2018, Pillepich2018b, Naiman2018, Marinacci2018, Nelson2019}.
%The work presented here investigates the contribution of galaxy mergers to the quenching of the star-forming galaxy population. 
%Thanks to large simulated volumes provided by recent cosmological hydrodynamical simulations, we can quantify the effects of galaxy mergers using statistical galaxy samples.
%The primary goal of this project is to assess the role played by galaxy mergers in building up the quenching population. 
%By taking advantage from the superb statistic offered by the IllustrisTNG simulations \citep{Springel2018, Nelson2018, Pillepich2018b, Naiman2018, Marinacci2018, Nelson2019}, 
We follow the evolution of simulated star-forming post-mergers within a realistic cosmological context. 
We then compare their evolution to that of non-post-merger galaxies with similar physical parameters and environment.

The paper is organised as follows: in Section~\ref{sec:method}, we introduce our methodology. In Section~\ref{sec:results}, we present the results quantifying the impact of galaxy mergers on star formation quenching. In Section~\ref{sec:bhmismatch} and Section~\ref{sec:reseffects}, we discuss the effects of our methodology on the results. Finally, we summarise our work in Section~\ref{sec:conclusions}.

\section{Data and Methods}
\label{sec:method}
\begin{figure*}
\centering
\includegraphics[width=\linewidth]{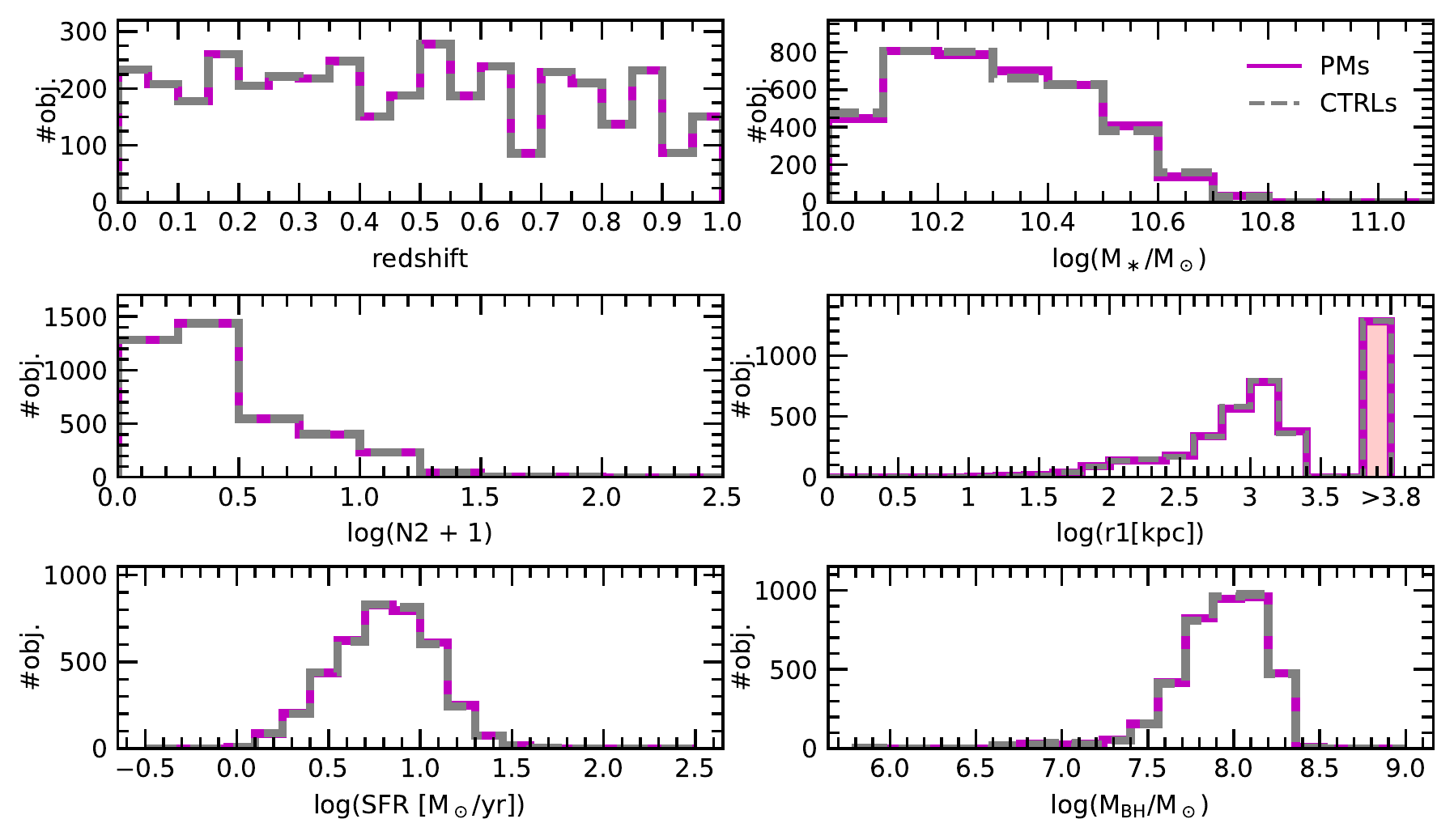}
\caption{A comparison between the distributions of redshift, M$_\ast$, $N_2$,  $r_1$, SFR, and  M$_\text{BH}$ of post-mergers (magenta) and their control galaxies (gray). The pink shaded bin in the r1 distribution includes all galaxies with $r_1 > 2$ Mpc. The control sample is very well matched to the post-merger one in all the six parameters. }
\label{fig:distribution}
\end{figure*}
\subsection{The IllustrisTNG simulation suite}
\label{sec:tng}
The work we present here is primarily aimed at quantifying the impact of galaxy mergers on star formation quenching.
We identify galaxy post-mergers in the IllustrisTNG simulation suite
\citep{Nelson2019} to study the relationship between mergers and the interruption of star formation within a cosmological framework.
The IllustrisTNG project includes a suite of large-box magnetohydrodynamical cosmological simulations in a $\Lambda$CDM Universe
which provides an exquisite sample of galaxies spanning a variety of galaxy properties (e.g., mass, environment, star formation rate -SFR-). 
Here we briefly summarise the main characteristics of the simulations. 
The simulations and physical model are introduced in detail in \citet{Marinacci2018, Naiman2018, Nelson2018, Pillepich2018b, Springel2018}.
IllustrisTNG (or TNG) is the descendant of the Illustris cosmological simulation \citep{Vogelsberger2013, Vogelsberger2014} with an improved physical models and numerical scheme. Moreover, TNG 
introduces a number of additional features to obtain a better agreement with observational results. 
In this paper, we focus on TNG300-1, the highest resolution run for the largest publicly released volume of $302^3$ cMpc$^3$. TNG300-1 offers the largest statistics, whilst still guaranteeing adequate numerical resolution. TNG300-1 has $2500^3$ initial resolution elements, with dark matter and stellar mass resolutions of $5.9\times10^7$M$_\odot$ and
$1.1\times10^7$M$_\odot$, respectively.
The simulation runs from redshift $127$ to the present day
using the AREPO moving-mesh code \citep{Springel2010, Pakmor2016}. The cosmological parameters used in IllustrisTNG are in accordance with \citet{PlanckCollaboration2016} which is given by a matter density $\Omega_\text{M,0} = 0.3089$, baryon density $\Omega_\text{b,0} = 0.0486$, dark energy density $\Omega_{\Lambda,\text{0}} = 0.6911$, and a Hubble parameter $h = 0.6774$. 
For our purposes, it is relevant to introduce some features of the IllustrisTNG physical model:
\begin{enumerate}
\item \emph{Star formation:} star formation
occurs in a pressurised, multi-phase interstellar medium following the \citet{Springel2003} formalism.
Gas particles whose density exceeds a threshold of $\sim0.1$ cm$^{-3}$ are ``star-forming'' and their gas is converted to stars stochastically following the Schmidt-Kennicutt law \citep{Kennicutt1998} assuming a \citet{Chabrier2003} initial mass function \citep[see][for further details]{Nelson2015, Pillepich2018a}.
\item \emph{Black holes and AGN feedback}:
black holes are seeded with an initial mass of $1.18\times10^6$M$_\odot$ at the centres of the potential wells of haloes exceeding a threshold mass of $7.38\times10^{10}$M$_\odot$.
Black holes can grow their mass either through (1)
accretion following the modify Bondi-Hoyle scheme, or (2) mergers with other black holes. 
AGN feedback is directly related to the accretion rate onto the central black holes ($\dot{\text{E}} \propto \dot{\text{M}}_{\text{BH}}\text{c}^2$).
At high accretion rates (i.e., quasar mode
feedback), thermal energy is returned to the black hole's environment, whereas at low accretion rates (i.e., radio mode feedback, or kinetic feedback), energy accumulates until it reaches an energy threshold, then mechanical energy is instantaneously released along a random direction into the gas around the black hole \citep[see][for further details]{Weinberger2017}.
\end{enumerate}

%For the purpose of this paper, for each galaxy and for its descendants we retrieve and use the following quantities from the TNG database:
In the work presented here, each galaxy is parametrised by:

\begin{itemize}

\item \emph{Galactic radius} (\rgal): we define a galaxy's radius to be twice the stellar half mass radius.
%We adopt the radius defined by twice the stellar half mass radius as the size of a TNG galaxy.  

\item \emph{Stellar mass} (M$_\ast$): the sum of the masses of all stellar particles contained within \rgal{} from a galaxy's centre (which is defined as the position at the minimum of the gravitational potential).

\item \emph{Gas mass} (M$_\text{gas}$): the gas mass is measured by summing the mass of all gas particles within \rgal{} from the galaxy's centre.

\item \emph{Black hole mass} (M$_\text{BH}$): is the mass of the supermassive black hole at the minimum of a galaxy's potential well. 

\item \emph{SFR}: this is the instantaneous star formation rate within \rgal{}. 
We use a metric based on the SFR to select and analyse the TNG post-merger galaxies.  
Specifically, we fit a redshift dependent star-forming main sequence (SFMS) to the TNG sample, and define quenched galaxies as those lying at least below $-3\sigma$ from the SFMS.

\item \emph{Cumulative kinetic feedback} ($\int \dot{\text{E}}_\text{kinetic} \text{dt}$): in our analysis we use the total amount of kinetic AGN feedback energy injected into the gas surrounding the central black hole in the low accretion rate mode, accumulated over the entire lifetime of the black hole in the centre of the galaxy. 
During black hole mergers the cumulative kinetic energy is summed for the two merging components. 

\item \emph{Potential} (V(r$_\text{gas}$)): the gravitational potential energy experienced at the position of a gas cell (r$_\text{gas}$), representing the energy required to unbind the gas. 
V(r$_\text{gas}$) is used to calculate the total gravitational binding energy of the gas particles as in \citet{Terrazas2020}:
\begin{equation}
\text{E}_\text{binding}(\leq \text{R}_\text{gal}) = \frac{1}{2}\sum_{\leq\text{R}_\text{gal}} \text{m}_\text{gas} \text{V}(\text{r}_\text{gas}), \label{eq:bind}
\end{equation}
where the sum is extended to all the gas particles  (m$_\text{gas}$) within \rgal{}. 
%The total binding energy defines the amount of energy that should be provided to the gas to prevail over the gravitational attraction that keeps the gas bound within the galaxy.
The total binding energy defines the amount of energy that is needed to void a galaxy of the gas within \rgal{}.
\end{itemize}

%%%%%%%%%%%%%%%%%%%%%%%%%%%%%%
\subsection{Star-forming post-mergers}
%%%%%%%%%%%%%%%%%%%%%%%%%%%%%%
%We select and follow the evolution of star-forming post-merger galaxies from the IllustrisTNG merger trees created using
We use the post-merger galaxy sample identified in \citet{Hani2020}, wherein galaxy mergers are defined as nodes in the {\scshape Sublink} merger trees \citep{RodriguezGomez2015}. 
% {\scshape Sublink} (Rodriguez-Gomez et al. 2015). 
Namely, we define a post-merger (or PM) in the snapshot immediately after the coalescence phase as the remnant of two interacting galaxies.
% (i.e. $\leq 162$ Myr after the merging, at the snapshot time resolution of the simulation).
Given the time resolution of the TNG snapshots, this approach identifies mergers within $\sim160$ Myr after coalescence (i.e. the average time between successive snapshots at $\text{z} < 1$).
%Among all the post-mergers in TNG 300-1, we restrict our analysis to those satisfying the following criteria:
%In the work presented here, we are interested in star-forming post-mergers unlike \citet{Hani2020} who include passive post-mergers. 
Following \citet{Hani2020}, our post-merger sample is restricted to those satisfying the following criteria:
\begin{itemize}

\item z $\leq 1$: we only follow the redshift evolution of post-mergers in the last $\sim8$ Gyr.

\item M$_\ast \geq 10^{10}$ M$_\odot$: in IllustrisTNG, galaxies are well resolved above M$_\ast>10^{9}$ M$_\odot$ (i.e. $\geq 90$ stellar particles per galaxy at the resolution of TNG 300-1). 
Therefore, our criterion ensures a complete sample of post-mergers with mass ratio  (secondary/primary) larger than 1:10. 

\item  The mass ratio  (secondary/primary, $\mu$) in the range $0.1 \leq \mu \leq 1$. 
 %This criterion is strictly connected with the previous one, and it prevents the analysis of remnants of mergers with a companion less massive than the reliable mass limit of the simulation. 
 We prevent numerical stripping issues by adopting the maximum stellar mass over the past $0.5$ Gyr for all mass ratio calculations following \citet{Patton2020}.
 
 \item The relative separation from the nearest neighbour r$_\text{sep} \geq 2$. The parameter r$_\text{sep}$ is defined by \citet{Patton2020} as:
 \begin{equation}
 \text{r}_\text{sep} = \frac{\text{r}}{\text{R}^\text{host}_\text{1/2} + \text{R}^\text{comp}_\text{1/2}},
 \end{equation}
where r is the 3D separation between the centres of the host (i.e. post-merger in our case) and its closest neighbour, and R$^\text{host}_\text{1/2}$ and R$^\text{comp}_\text{1/2}$ are the stellar half mass radii of the post-merger and the closest neighbour, respectively. 
The criterion r$_\text{sep} \geq 2$ avoids post-mergers that are undergoing further close interactions which could interfere with the descendant's evolution. 

\item The time elapsed since the previous merger must be larger than $2$ Gyr, i.e. we exclude post-mergers that have experienced another merger within the last $2$ Gyr.
This criterion reinforces our purpose of isolating the effect of a single merger on galaxy evolution.

\item Galaxies must be star-forming when first selected (in order that we can later observe them quenching). 
In practice this is implemented by requiring SFRs higher than $-1\sigma$ from the star-forming main sequence best-fit.
%logarithmic offset from the star-forming main sequence ($\Delta$SFR) $\geq -0.2$ dex. 
Unlike the original sample by \citet{Hani2020} who include passive post-mergers, in the work presented here, we are interested in star-forming post-mergers in order that we may track whether the merger causes them to quench.
%We stress that demanding only star-forming post-mergers limits the stellar mass range to a maximum of $10^{10.8}$ M$_\odot$.  
Given the arbitrary choice of our SFR threshold value, we investigate the impact of imposing different limits to separate quenched and star-forming galaxies in Section~\ref{sec:results}.
%In the SFR-M$_\ast$ plane, at each snapshot (i.e. each redshift sampled in the simulation) we define the star-forming main sequence (SFMS) as the linear best-fit of the star-forming galaxies (e.g. Brinchmann et al. 2004). Then, for each galaxy, we evaluate the relative importance of star formation by 
%calculating, on a logarithmic scale in the SFR-M$_\ast$ plane, the vertical offset between its SFR and the SFMS (hereafter, $\Delta$SFR).
%Since we find that the scatter around the SFMS is of about $0.3$ dex at any snapshot (in accordance with observational results), we define as quenching galaxies those which show a $\Delta$SFR$<-0.9$ dex (i.e. below $-3\sigma$ from the SFMS). In TNG this threshold corresponds to a specific SFR (i.e. log$_{10}$ SFR/M$_\ast$) of about $-11$ yr~$^{-1}$, a limit often used in observational studies to separate passive galaxies from star-forming ones (e.g. Ilbert et al. 2010,2013; Pozzetti et al. 2010).
%Nevertheless, given the arbitrary choice of our threshold value, we investigate the impact of imposing different limits to separate quenching from star-forming galaxies. 
%We re-run our analysis by applying both less conservative thresholds (e.g.$-2\sigma$ from the SFMS) and more conservative ones (e.g.  $-4\sigma$ and $-5\sigma$ from the SFMS), and we do not find a significant difference in the main result we present in the next sections of this paper. 
\end{itemize}

\noindent Our selection criteria yield a sample of $3472$ star-forming post-mergers with stellar masses   $10^{10} \leq$ M$_\ast/$M$_\odot \leq 10^{10.8}$, at redshift z $\leq 1$. We note that $\lesssim10\%$ of post-mergers in our sample have a mass ratio $\mu > 0.5$, therefore the results we present in this paper are statistically dominated by minor mergers ($\mu < 0.3$).

\subsection{Statistical control sample}
\label{sec:control_sample}
We are interested in investigating the link between galaxy mergers and the quenching of star formation. 
We quantify the impact of galaxy mergers on the quenching of star-formation using an observational approach that consists of identifying control galaxies that are matched to each post-merger galaxy in redshift, stellar mass, local density, and isolation \citep[e.g.,][]{Ellison2013, Patton2013}. 
In this section, we describe the steps of creating the control sample.

We implement an adaptation of the matching procedure used in \citet{Patton2016} and \citet{Patton2020} to select statistical controls (CTRL) for galaxies in our sample \citep[see also][]{Hani2020}.   
For each star-forming post-merger, we first define a star-forming control pool (i.e. non-post-merger galaxies with SFR higher than $-1\sigma$ from the SFMS), in the same snapshot (i.e. same redshift), with M$_\ast \geq 10^{10}$ M$_\odot$, and a relative separation from the nearest neighbour r$_\text{sep} \geq 2$. 
We then reject galaxies that have experienced a merger ($\mu \geq 0.1$) within 2 Gyr (i.e. t$_\text{PM} \geq 2$ Gyr). 
%Moreover, we narrow the control sub-sample of each post-merger to the galaxies with a match in:
Then, for each post-merger in our sample, we identify the control galaxies that match the post-merger properties as follows:  
\begin{itemize}
\item log(M$_\ast$) within a tolerance of $0.05$ dex.

\item The environmental parameters $N_2$ and $r_1$, defined by \citet{Patton2016} as the number of neighbours within a radius of $2$ Mpc (i.e. local density), and the distance to the nearest neighbour with M$_\ast \geq 0.1 \times $M$_{\ast,\text{host}}$ (i.e. isolation), respectively, within a tolerance of 10\%.

\item log(SFR) within a tolerance of $0.01$ dex. 

\item log(M$_\text{BH}$) within a tolerance of $0.05$ dex. By matching in black hole mass we prevent possible bias related to the AGN feedback model (see Section~\ref{sec:AGN}). 
In Section~\ref{sec:bhmismatch}, we discuss how this requirement affects our results.
%As mentioned in section \ref{sec:AGNmodel}, above a threshold mass of M$_\text{BH} \sim 10^{8.2}$ M$_\odot$, the momentum injected into the medium surrounding the black hole by kinetic AGN feedback produces outflows of gas, that eventually will quench the galaxy. 

\end{itemize}
If more than one control is found for a given post-merger, we follow the weighting scheme of \citet{Patton2016} to select up to $5$ control galaxies. 
We then define the post-merger's \textit{control galaxy} as the single best control galaxy that shares the most number of subsequent snapshots with the post-merger's descendants whilst maintaining environmental parameters within $40\%$ of the descendants'. 
%In the $\sim 97\%$ of cases we found at least one control galaxy for every star-forming PM. In the other cases, we increase the tolerance range (and, if required, up to a maximum of) by 0.05 (0.1) dex for the M$_\ast$, 0.01 (0.04) dex for the SFR, 0.025 (0.1) dex for the M$_\text{BH}$ and 10\% (20\%) for both $N_2$ and $r_1$. 
%If more than one control is found, we follow the weighting scheme of Patton et al. (2016) to select up to five best matches in all the parameters. 
%In these cases, we choose the control galaxy to be the one among the best five that has the maximum number of subsequent snapshots in common with the post-merger, and that keeps a match with the post-mergers descendants in the environmental parameters $N_2$ and $r_1$, within a tolerance of 40\%. 
\autoref{fig:distribution} shows the distributions of redshift, M$_\ast$, $N_2$,  $r_1$, SFR, and  M$_\text{BH}$ for the star-forming post-mergers and their control galaxies. 
The matching process offers a control population that well matches our post-merger sample in all the aforementioned parameters.

\begin{figure}
\centering
\includegraphics[width=\columnwidth]{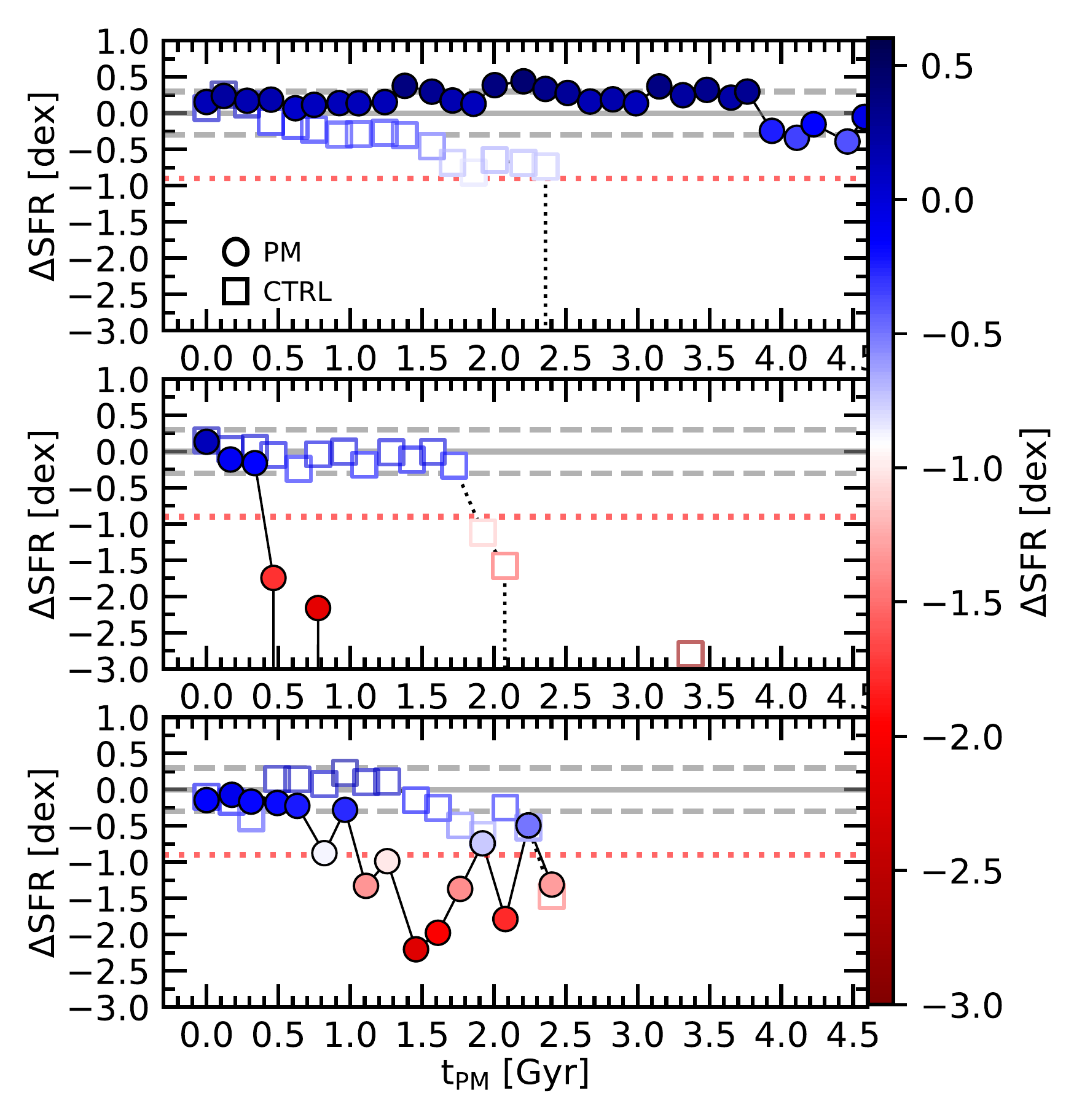}
\caption{Examples of the evolution of $\Delta$SFR after the merger  for three example star-forming post-mergers (filled symbols) and their controls (open symbols). 
The colour shading represents the $\Delta$SFR: shades of blue for $\Delta$SFR $\geq -0.9$, and reds for quenched descendants with  $\Delta$SFR $< -0.9$.
The two dashed grey lines represents the  scatter ($\Delta$SFR $\pm  0.3$ dex, i.e. $\pm 1\sigma$) around the SFMS (solid grey line). The red line ($\Delta$SFR $= -0.9$ dex, i.e. $-3\sigma$ below the SFMS) represents the threshold that separates star-forming (above the line) and quenched (below the line) galaxies.
The top panel shows a post-merger that remains star-forming after coalescence, while its matched control galaxy permanently quenches around $2.4$ Gyr after \tpm{}$=0$, as its $\Delta$SFR stays steadily below $-3$ dex until redshift z $=0$. The central panel shows a post-merger that quenches within $500$ Myr after the merger, while its matched control galaxy quenches around $2$ Gyr after \tpm{}$=0$.
Finally, the bottom panel depicts a post-merger that remains star-forming up to $1$ Gyr after the merger, then quenches, though not permanently (i.e. it experience some episodes of low SFR).  
The control galaxy, instead, keeps forming stars for the entire period over which we follow its evolution (i.e. at $\text{z}=0$).
The post-merger and control galaxy evolution depicted in the bottom panel end at $\sim2.5$ Gyr after $\text{t}_\text{PM}=0$, when the simulation reaches redshift $\text{z}=0$. }
\label{fig:example}
\end{figure}
Once the control galaxy sample has been identified, we follow the evolution of the SFR in descendants of  %with the time elapsed from the coalescence of the $\Delta$SFR of both post-mergers descendants and their controls.
post-mergers and controls forward in time through the simulation. 
For each galaxy, we evaluate the relative relevance of star formation by 
calculating, on a logarithmic scale in the SFR-M$_\ast$ plane, the vertical offset between its SFR and the SFMS (hereafter, $\Delta$SFR).
Then, we define galaxies to be quenched when their $\Delta$SFR drops below $-0.9$ dex (i.e., a deviation of $<-3 \sigma$ from the SFMS).
\autoref{fig:example} shows the evolution of $\Delta$SFR as a function of time since merger (t$_\text{PM}$) for three example post-mergers and their controls\footnote{When we express the time evolution of control galaxies in terms of \tpm{}, we measure the time relative to their matched post-mergers.}.
The top panel shows a post-merger that remains star-forming after coalescence, while its matched control galaxy quenches around $2.4$ Gyr after \tpm{}$=0$. The central panel shows a post-merger that quenches within $500$ Myr after the merger.
Finally, the bottom panel depicts a post-merger that remains star-forming up to $1$ Gyr after the merger, then quenches, though not permanently (i.e. it experience some episodes of low SFR).  
The control galaxy, instead, keeps forming stars for the entire period over which we follow its evolution (i.e. at $\text{z}=0$).

%At each snapshot, we compare the SFRs of post-mergers to their controls using the following metric:
We quantify the effects of mergers on star formation quenching using two metrics: (i) the fraction of post-merger descendants with quenched star formation, and (ii) the number of quenched post-mergers (\#\QPM{}) normalized by the number of quenched controls (\#\QCTRL{}) in $160$ Myr intervals (Q$_\text{excess}$):
\begin{equation}
\text{Q}_\text{excess} = \frac{\#\text{Q}_\text{PM}}{\#\text{Q}_\text{CTRL}}. \label{eq:excess}
\end{equation}
Q$_\text{excess}$ measures the relative tendency of mergers to experience quenching compared to other evolutionary processes which are accounted for in the controls. 
Taken together, these two metrics allow us to quantify both the absolute rate of quenching in post-mergers, as well as assessing whether quenching happens more frequently in post-mergers that in controls.  
%We classify as quenching galaxy those descendants whose $\Delta$SFR drops below $-0.9$ (i.e. $-3\sigma$ from the SFMS at that redshift in TNG300-1). 
%We analyse the relationship between merging and quenching in two alternatives ways. 
%The first one involves the evolution of the percentage of post-merger descendants that quench the star formation as a function of the time elapsed from the coalescence. 
%The second way concerns a comparison with the number of quenching galaxies predicted by similar but non-post-merger galaxies. 
%We count the number of quenching post-mergers and quenching controls in bins of $200$ Myr of time passed from the merging (t$_\text{PM}$), and we can define a simple metric
%\begin{equation}
%\text{Q}_\text{excess} = \frac{\#\text{Q}_\text{PM}}{\#\text{Q}_\text{CTRL}} (\text{t}_\text{PM}),
%\end{equation}
%that quantifies the tendency of a merging to quench the star formation with respect to the expected number for a non-post-merger population.  
%This is viable because the only difference in the evolution of post-mergers and control galaxies is the merging, therefore, we expect that an excess (or a lack) of quenching descendants could be associated only with the recent merging of the analysed post-mergers.

%%%%%%%%%%%%%%%%%%%%%%%%%%%%%%%%%%%%%%%%%%%%%%%%%%%%%
\section{Results and discussion}
\label{sec:results}
%%%%%%%%%%%%%%%%%%%%%%%%%%%%%%%%%%%%%%%%%%%%%%%%%%%%%
%%%%%%%%%%%%%%%%%%%%%%%%%%%%%%%%%%%%%%%%%%%%%%%%%%%%%
\subsection{Quenching in descendants of star-forming post-mergers}
\label{sec:ratio}
%%%%%%%%%%%%%%%%%%%%%%%%%%%%%%%%%%%%%%%%%%%%%%%%%%%%%
In this section, we analyse the impact of mergers on the interruption of star formation during the early stages after the coalescence phase. 
We apply the methods defined in Section~\ref{sec:control_sample} to quantify differences between the evolution of star-forming post-mergers and matched control galaxies that have not experienced any merger in the past $2$ Gyr.
\begin{figure}
\centering
\includegraphics[width=\linewidth]{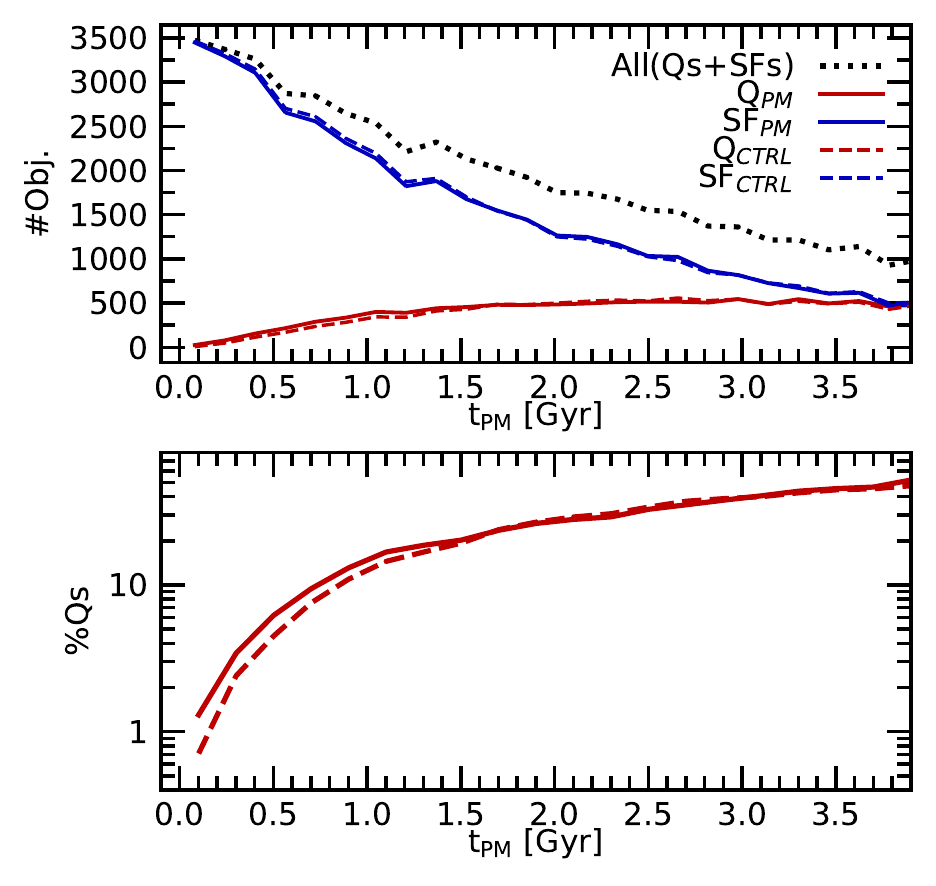}
\caption{(Top) The black dotted curve represents the total number of post-mergers (and an equal number of control galaxies) as a function of time elapsed since coalescence (t$_\text{PM}$); the solid curves represent the evolution of the number of post-mergers that are still star-forming (\SFPM{}, solid blue) or quenched (\QPM{}, solid red), respectively. The dashed curves represent the evolution of the number of still star-forming control galaxies (\SFCTRL{}, dashed blue) and quenched controls (\QCTRL{}, dashed red), respectively. 
(Bottom) The percentage of \QPM{} galaxies (solid red curve) and of \QCTRL{} galaxies (dashed red curve) as a function of t$_\text{PM}$. 
%The insets in both panels show a zoom of the first $1$Gyr from the merging. }
The figure demonstrates that the absolute fraction of quenched post-mergers is small shortly after coalescence, indicating that the merger does not promptly truncate star formation.
}
\label{fig:percent_evolution}
\end{figure}
The top panel in \autoref{fig:percent_evolution} shows the evolution of the total number of post-mergers and their associated controls as a function of time elapsed since coalescence (t$_\text{PM}$), in bins of $160$ Myr.
The black dotted curve represents the total number of post-mergers (and, by definition, an equal number of control galaxies) in each bin. The total number of galaxies decreases with time for three reasons: (1) the merger rate increases with increasing redshift \citep[e.g.][]{Lin2008, Lotz2011}, (2) we analyse the post-mergers in the range $0\leq\text{z}\leq1$, therefore the maximum time after the merger that we can analyse decreases as the snapshots approach $\text{z}=0$, and (3)  we only follow the evolution of a given post-merger up to the moment when its descendant, or the descendant of its control galaxy, experiences a new close encounter or a merger. 
We also interrupt tracing the evolution when the match between the environmental parameters (i.e. N2 and r1) exceeds a tolerance of $40\%$. 
%We split the total number of post-mergers in each bin into star-forming and quenched post-mergers.
We divide the post-merger population, at any given time, into those that are star-forming and those that are quenched (recalling that we require the post-mergers to be star-forming at t$_\text{PM}=0$).
\autoref{fig:percent_evolution} shows that, in the early phase after the merger, the vast majority of post-mergers remain star-forming, with only a small fraction of them becoming quenched (e.g. $\sim 50/3500$ at \tpm{} $= 125$ Myr). 
%With increasing time since coalescence, the number of quenched post-mergers increases slightly, and stabilises around $400$ at \tpm{} $= 1$ Gyr. 
The control sample shows a qualitatively similar behaviour but with an even smaller number of quenched systems at \tpm{} $< 1$ Gyr. 
%An alternative way to show this trend is offered by the evolution of the percentage of quenching post-mergers with the time after the coalescence (see bottom panel of \autoref{fig:percent_evolution}). 
The bottom panel of \autoref{fig:percent_evolution} provides a complementary perspective by showing these results as percentages of each population. 
We find that only $1.4\%$ of the star-forming post-mergers quench their star formation within $125$ Myr following coalescence, compared to $0.7\%$ for the controls. 
The fraction of quenched post-mergers rises to $\sim5\%$ at \tpm{} $= 500$ Myr ($3.5\%$ for the controls) and $16\%$ within $1$ Gyr from the merger ($12.5\%$for the controls).  
After $1.5$ Gyr the fraction of quenched post-mergers and control galaxies grow together with no appreciable distinction. 
We conclude from \autoref{fig:percent_evolution} that the process of coalescence does not result in widespread quenching of the post-merger population.
%The rarity of quenching in post-mergers is qualitatively in accordance with other cosmological simulations and observational results. For instance, 
%\citet{RodriguezMontero2019} find that major mergers in the SIMBA simulation \citep{Dave2019} are not directly related to quenching, as the typical delay between the merger and subsequent quenching is larger than $1$ Gyr.
%By analysing a sample of galaxies extracted from the SDSS DR7, \citet{Weigel2017} show that major mergers should not be the preferred path leading to permanent quenching. They found that major merger quenched galaxies account for a maximum of $5$\% of the quenched population at a given stellar mass, both at low- and intermediate-redshift. 
\begin{figure}
\centering
\includegraphics[width=\linewidth]{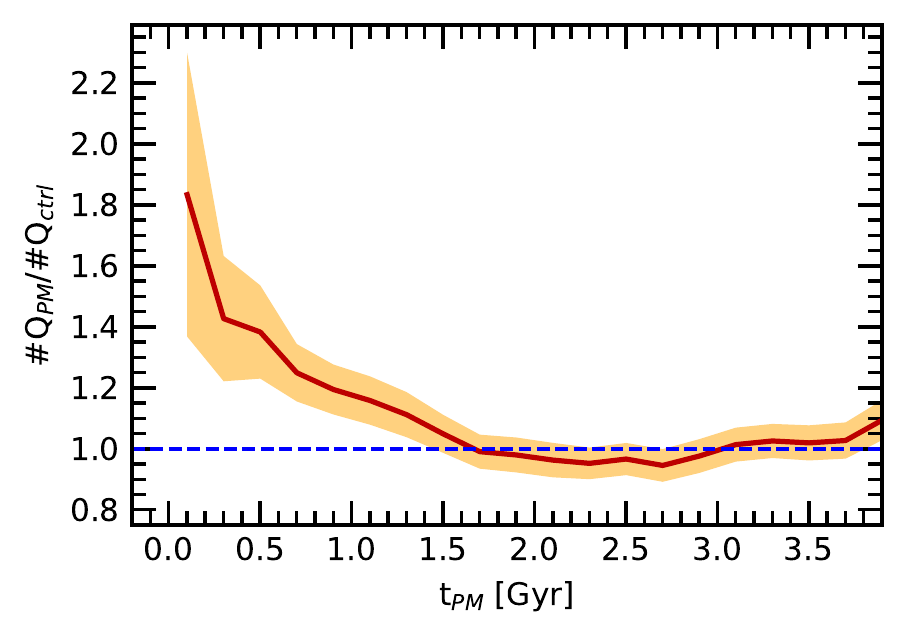}
\caption{The ratio between the number of quenched post-mergers and quenched control galaxies as a function of the time elapsed since coalescence t$_\text{PM}$. The shaded contours represent the error, that is quantified as the propagation of the Poissonian $1\sigma$ error of the ratio.
The figure demonstrates that post-mergers show an excess of quenched galaxies, relative to expectations from the control of up to a factor $\sim2$.  
The excess of quenched galaxies persists for approximately $1.5$ Gyr after the merger.}
\label{fig:ratio_evolution}
\end{figure}
%We find that quenching in the star-forming post-merger population is rare in TNG300-1. 
%Although, the trends of post-mergers and control galaxies appear to be slightly different at \tpm{} $< 1.5$ Gyr. While control galaxies follow an isolated evolution, the difference could reveal subtle alterations in the evolution of the descendants as a result of the merging.
Nonetheless, the quenched fraction is slightly larger (an effect we quantify shortly) in the post-merger sample compared to the controls at \tpm{} $\leq1.5$ Gyr, which may be indicative of the \textit{subtle} effects of galaxy mergers on the quenching of merger descendants, which provide a low level facilitation of the quenching process.  

In \autoref{fig:ratio_evolution}, we quantify 
%this discrepancy between the two populations by showing the ratio between the number of quenching post-mergers and quenching control galaxies as a function of \tpm{}.
the enhancement in quenched post-mergers relative to their controls as the ratio of quenched post-mergers to quenched control galaxies. 
It is important to stress that t$_\text{PM} = 0$ is not displayed in \autoref{fig:ratio_evolution}. By construction, we selected star-forming post-mergers and star-forming control galaxies, therefore there are $0$ quenched post-mergers and control galaxies at t$_\text{PM} = 0$. 
Around $150$ Myr after coalescence, we find that post-mergers are quenched with an excess of $1.83\pm 0.47$ compared to their controls which have not experienced a recent merger (within the past $2$ Gyr). The error is quantified as the propagation of the Poissonian $1\sigma$ error of the ratio in  \autoref{eq:excess}.
The excess is confirmed at a significance level of $1.8\sigma,$ with respect to $\#$\QPM$/\#$\QCTRL{}$=1$. 
We find that	the excess is larger in the early post-merger phase when the mergers' effects are stronger \citep[e.g.,][]{Hani2020}, and then it decreases steadily towards $\#$\QPM$/\#$\QCTRL{}$=1$. Beyond $1.5$ Gyr after the merger, the number of quenched post-mergers becomes statistically indistinguishable from that of the control sample. 
%We interpret the temporariness of the excess as though mergers are able to accelerate the process leading to quenched star formation galaxies whose progenitor had physical conditions close to sustaining effective AGN kinetic feedback.
We speculate that this short term enhancement of quenched post-mergers is the result of the merger expediting quenching in a system that was already close to achieving the conditions necessary for halting star formation (i.e. a critical AGN feedback).  
We return to this in Section~\ref{sec:AGN}.

In order to test whether our results presented in Figures~\ref{fig:percent_evolution} and \ref{fig:ratio_evolution} depend on our definition of quenched and star-forming galaxies, we investigate the impact of different $\Delta$SFR thresholds. 
%In the SFR-M$_\ast$ plane, at each snapshot (i.e. each redshift sampled in the simulation) we define the star-forming main sequence (SFMS) as the linear best-fit of the star-forming galaxies (e.g. Brinchmann et al. 2004). Then, for each galaxy, we evaluate the relative importance of star formation by calculating, on a logarithmic scale in the SFR-M$_\ast$ plane, the vertical offset between its SFR and the SFMS (hereafter, $\Delta$SFR).
Recall that since we find that the scatter around the SFMS is of $\sim0.3$ dex at any given snapshot (in accordance with observational results), our fiducial threshold for labelling a galaxy as quenched is $\Delta\text{SFR}<-0.9$ to be quenched  (i.e. below $-3\sigma$ from the SFMS). In TNG this threshold corresponds to a specific SFR (i.e. log$_{10}$ SFR/M$_\ast$) of about $-11$ yr$^{-1}$, a limit often used in observational studies to separate passive galaxies from star-forming ones \citep[e.g.,][]{Ilbert2010, Ilbert2013, Pozzetti2010}.
We re-run our analysis by applying both less conservative thresholds ($-2\sigma$ from the SFMS) and more conservative ones ($-4\sigma$ and $-5\sigma$ from the SFMS), and we do not find a significant difference in the main result presented in this section.

In the next two sub-sections, we will explore the properties of the post-mergers that do/do not quench.  
In order to distinguish a quenching event that can plausibly be linked directly to the merger, we re-define the quenched post-merger sample as those galaxies which quench within $0.5$ Gyr after coalescence ($171$ galaxies).  I.e. we remove from the \QPM{} sample the $589$ post-mergers that quench on timescales longer than $0.5$ Gyr.  
Since none of the control galaxies have experienced a significant merger within at least $2$ Gyrs (by definition), all of the quenched controls are retained in the \QCTRL{} sample ($772$ galaxies).  Thus, the \QPM{} and \QCTRL{} samples represent galaxies that have/have not quenched as the result of a recent merger.  The star-forming samples (\SFPM{} and \SFCTRL{}) also remain unchanged and contain $2114$ and $2102$ galaxies, respectively.  These are the samples used in the following sub-sections.

%However, given the different criteria adopted to define them, we need to make some assumptions to guarantee a fair comparison. 

 %adopted to select the \QCTRL{} galaxies,  for this sub-sample the \autoref{eq:percch} is measured between the time of their quenching and the amount of gas 

%in the following we analyse the properties of the \QPM{} in the snapshot where they are quenched (

%against those of the the star-forming sub-samples 

%To allow a comparison of gas percentage range of \QPM{} with the other three sub-samples, we need to make some assumptions \QCTRL{} populations, we  

%%%%%%%%%%%%%%%%%%%%%%%%%%%%%%%%%%%%%%%%%%%%%%%%%%%%%
\subsection{Gas evolution in TNG post-mergers and control galaxies}
\label{sec:gasevol}
%%%%%%%%%%%%%%%%%%%%%%%%%%%%%%%%%%%%%%%%%%%%%%%%%%%%%
%The physical mechanism provoking the interruption of star formation could act by altering the gas properties and distribution within star-forming galaxies. 
Altering the absolute gas content, or its spatial distribution, is a possible mechanism for driving quenching.
In the literature, processes that remove gas from the reservoir of star-forming galaxies are described as \emph{ejective} feedback, while those that slow down the cooling of gas or halt gas inflow from the halos surrounding galaxies are described as a \emph{preventive} feedback \citep[e.g.,][]{Somerville2015}. 
%However, quenching mechanisms can include a combination of both types of feedback. 
In practice, quenching could result from both types of feedback.
In this section, we explore the evolution of the amount of gas in the reservoir of TNG post-mergers, to understand what kind of feedback leads to the quenching of star-formation.
%Moreover, with this analysis we might find some clues about two different quenching processes acting in \QPM{} and \QCTRL{} populations; therefore, we could explain the excess of quenching post-mergers with respect to their control galaxies. 
Moreover, we explore whether the process(es) that drive quenching are the same in the quenched post-mergers (i.e. \QPM{}) and quenched control galaxies (i.e. \QCTRL{}).
\begin{figure}
\centering
	\includegraphics[width=\columnwidth]{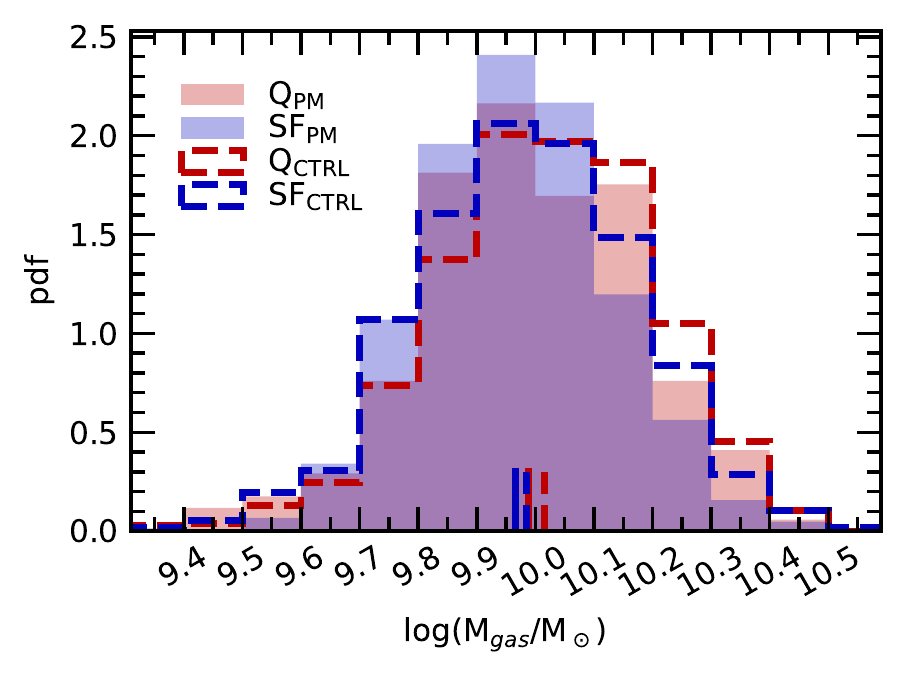}
	\caption{Distributions of the gas mass at $\text{t}_\text{PM}=0$. The post-mergers are represented by filled histograms: red for post-mergers that will quench within $500$ Myr after coalescence (\QPM{} galaxies), and blue for post-mergers that keep forming stars at least $500$ Myr after coalescence (\SFPM{} galaxies). Control galaxies are represented by the dashed open histograms: red for control galaxies that will quench (\QCTRL{} galaxies), and blue for control galaxies that keep forming stars ( \SFCTRL{}). 
	%The solid gray straight line represents the median, and the two dashed gray straight lines represent the $16^\text{th}$ and the $84^\text{th}$    percentiles of the distribution.}
	The vertical coloured ticks around the bottom/centre of the distributions indicate the median of the respective distributions.
	The figure shows that, at $\text{t}_\text{PM}=0$, our four populations (\QPM{}, \SFPM{}, \QCTRL{}, and \SFCTRL{}) have very similar distributions of gas.
	% between $10^{9.4} \leq \text{M}_\text{gas}/\text{M}_\odot \leq 10^{10.5}$ , with a median at roughly M$_\text{gas} = 10^{10}$ M$_\odot$.}
	}
    \label{fig:initial_gas}
\end{figure}

To analyse the nature of the feedback responsible for quenching in post-merger galaxies, we start by studying the evolution of gas in quenched post-merger galaxies and in the other three samples.
In \autoref{fig:initial_gas} we show the distribution of the amount of gas at $\text{t}_\text{PM}=0$ in quenched post-mergers (\QPM{}), quenched control galaxies (\QCTRL{}), star-forming post-mergers (\SFPM{}) and  star-forming control galaxies (\SFCTRL{}). 
At \tpm{} $=0$, the four sub-samples have similar gas masses between $10^{9.4}$\msol{} and $10^{10.55}$\msol{}, with over $50\%$ of galaxies possessing a gas mass larger than $10^{9.95}$\msol{}.
%The Kolmogorov-Smirnov test confirms, at a significance level $\alpha = 0.05$, that the \QPM{} population has the same initial gas mass distribution as the \SFCTRL{} and \QCTRL{}. Instead, the test reveals that \SFPM{} galaxies are drawn from a different population. 
%However, the difference is marginal, and it appears at gas masses larger than $10^{10}$\msol{}, where we find $40\%$ of \SFPM{}, against $\sim35\%$ of the other sub-samples. 
\begin{figure}
\centering
	\includegraphics[width=\columnwidth]{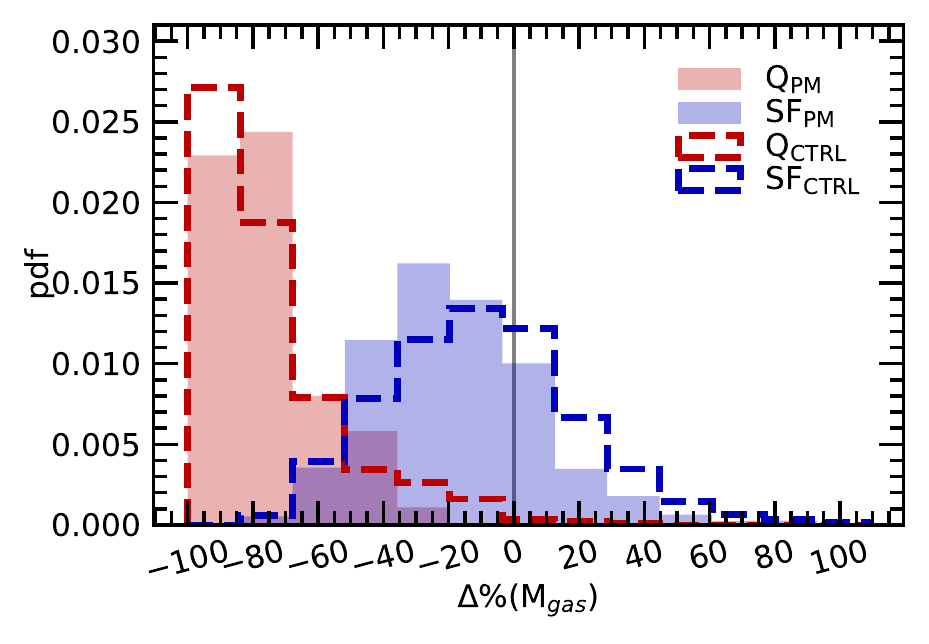}
	\caption{Distributions of the gas percentage change in galaxies ($\Delta\%[\text{M}_\text{gas}]$). $\Delta\%[\text{M}_\text{gas}]$ represents the percentage of gas lost with respect to the gas distribution shown in \autoref{fig:initial_gas}. The post-mergers are represented by filled histograms: red for quenched post-mergers (\QPM{} galaxies), and blue for star-forming post-mergers (\SFPM{} galaxies). Control galaxies are represented by the dashed open histograms: red for quenched control galaxies (\QCTRL{} galaxies), and blue for star-forming control galaxies (\SFCTRL{}). 
	The vertical gray line separates galaxies with a net gas loss (left) from those with a net gas gain (right).
	The figure demonstrates that quenching in post-mergers is characterized by a large fractional gas loss, and that there is neither a substantial difference between gas loss in \QPM{} and \QCTRL{} galaxies, nor between \SFPM{} and \SFCTRL{} galaxies.}
    \label{fig:gasloss}
\end{figure}

We quantify the fractional change in gas mass between t$_\text{PM}=0$ (or t$_0$) and the time at which a galaxy quenches (t$_\text{Q}$) as follows:
\begin{equation}
\Delta\%[\text{M}_\text{gas}] = \frac{\text{M}_\text{gas}(\text{t}_\text{Q}) - \text{M}_\text{gas}(\text{t}_0)}{|\text{M}_\text{gas}(\text{t}_0)|} \times 100. \label{eq:percch}
\end{equation}
Since we select the \QCTRL{} sub-sample without any constraints on the time of the quenching (see Section~\ref{sec:ratio}), several \QCTRL{} galaxies quench on very long ($\gtrsim \text{Gyr}$) timescales. 
Therefore, we calculate $\Delta\%[\text{M}_\text{gas}]$ for the \QCTRL{} sample relative to their gas mass $\sim500$Myr before they quench which is consistent with the timescale used for comparison for the post-merger sample.
For the \SFPM{} (and \SFCTRL{}), we instead evaluate $\Delta\%[\text{M}_\text{gas}]$ at the third snapshot after coalescence (i.e. $\sim500$Myr).
The red coloured histogram in \autoref{fig:gasloss} shows the $\Delta\%[\text{M}_\text{gas}]$ distributions for the quenched post-merger galaxies (\QPM{}). 
We find that all \QPM{} galaxies have at least $25\%$ less gas than their initial amount. However, $68\%$ of them (between $16^\text{th}$ and $84^\text{th}$ percentile) experienced a much larger gas removal in the range between $60-92\%$ and a median loss of $82\%$ of the initial gas mass. 
%In the same figure, we can compare this result with the profile of \QCTRL{} galaxies. The curves of the two quenching populations appear to be slightly different only in the low-hand tail of the distribution below the $20^\text{th}$ percentile, while there is an almost perfect match above it. 
The distribution of quenched post-merger galaxies shows remarkable similarity with that of the quenched control galaxies sample (red dashed histogram), thus suggesting that quenching is not special in post-mergers, and that it is characterized by gas loss in both post-mergers and controls.
\autoref{fig:gasloss} also shows the gas percentage change of the star-forming post-merger population (\SFPM{}, blue coloured histogram) and star-forming control population (\SFCTRL{}, blue dashed histogram). 
These two star-forming sub-samples show distributions of $\Delta\%[\text{M}_\text{gas}]$ different from those of the quenched galaxies: the median values for \SFPM{} and \SFCTRL{} galaxies are $\Delta\%[\text{M}_\text{gas}] = -25\%$  and $-10\%$, respectively, and almost $25\%$ of \SFPM{} and $35\%$ of  \SFCTRL{} galaxies have even accreted gas mass ($\Delta\%[\text{M}_\text{gas}]>0$) with respect to the initial amounts.  

Possible reasons for the decline in the measured gas mass in galaxies include: (1) conversion of gas into stars, 
%(2) an illusory gas loss due to a decreasing \rgal{} that results in including fewer gas particles within the new galactic radius (we remind the reader that \rgal{} is defined as twice the radius containing half of the stellar mass and all the parameters we use in the analysis, including M$_\text{gas}$, are measured within \rgal{}), and 
and (2) the gas has been moved beyond \rgal{}. 
Considering the case of gas conversion into stars, this would require converting $\sim8\times10^9$\msol{} of gas into stars in about $500$ Myr, equivalent to a steady SFR$\geq 15$ \msol/yr. 
Although post-merger galaxies show particularly enhanced star formation activity, their star-burst phase lasts for only some tens Myr after coalescence and then their SFR drops exponentially to a more typical star-forming level \citep[e.g.,][]{Hani2020}.  
\begin{figure}
\centering
	\includegraphics[width=\columnwidth]{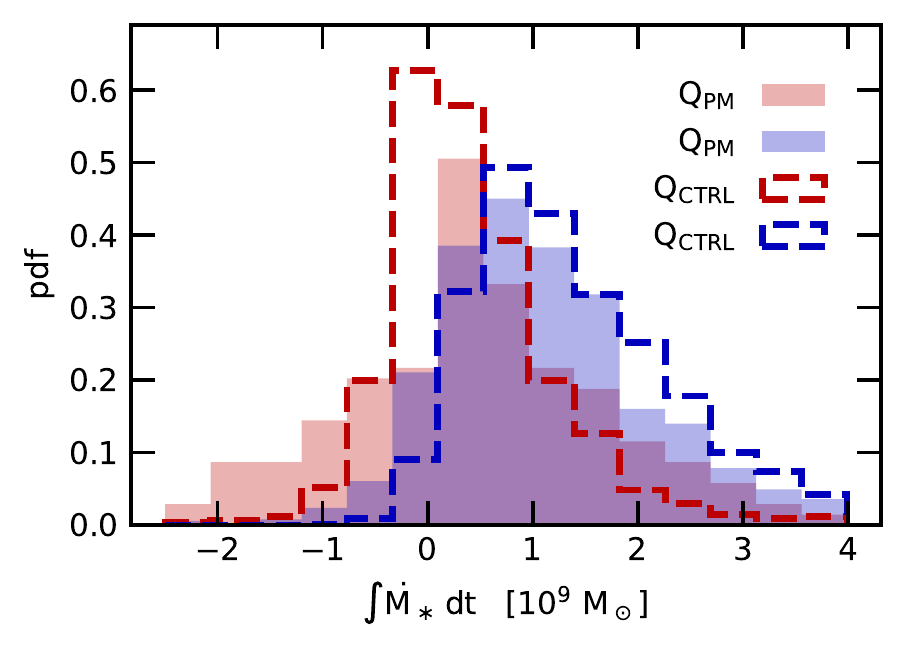}
	\caption{Distributions of stellar mass growth measured in the same interval of time as for \autoref{fig:gasloss}.
The post-mergers are represented by filled histograms: red for quenched post-mergers (\QPM{} galaxies), and blue for star-forming post-mergers (\SFPM{} galaxies). Control galaxies are represented by the dashed open histograms: red for quenched control galaxies (\QCTRL{} galaxies), and blue for star-forming control galaxies (\SFCTRL{}). 
%	The relation between the variation in \rgal{} and the growth in as a function of stellar mass. The post-mergers are represented by filled circles (red symbols for post-mergers that have quenched within $0.5$ Gyr from the merging, whereas blue symbols for post-mergers descendants that remain star-forming). Instead, the control galaxies are represented by open squares (red symbols for control galaxies that have quenched within $0.5$ Gyr from \tpm{}, whereas blue symbols for control galaxies descendants that remains star-forming). The black errorbars represent the$16^\text{th}-84^\text{th}$ percentile interval of the distribution of objects on that parameter.}
%%
%The figure reveals that the conversion of gas into stars in \QPM{} and \QCTRL{} galaxies is $8-10$ times less than the amount expected if the gas was being converted into stars, whilst the stellar mass growth in  \SFPM{} and \SFCTRL{} galaxies could be consistent with their gas loss. 
}
    \label{fig:RvsM}
\end{figure}
\autoref{fig:RvsM} shows the stellar mass growth distributions in our four sub-samples, measured in the same interval of time as for \autoref{fig:gasloss}. 
In the interval of time between the merger and quenching, the stellar mass in quenched post-merger galaxies increases, on average, by $\sim10^{8.7}$\msol{}, with a maximum of $\sim10^{9.3}$\msol{}. 
%Such an amount of new stars' mass is far below the measured gas mass loss.  
This amount of new stellar mass is far lower the measured loss in gas mass.
For example, we find that quenched post-mergers that show an increment in stellar mass of $\sim10^{8.7}$\msol{}, have experienced a gas loss in the range $10^{9.5}-10^{10.1}$\msol{} (i.e. $16^\text{th}-84^\text{th}$ percentiles), with a median gas loss of $10^{9.8}$\msol{}.

We do not find a significant difference between quenched post-mergers and quenched control galaxies. 
%About the variation of \rgal{}, we find that the bulk of \QPM{} and \QCTRL{} galaxies experience a marginal increase in \rgal{}.  
%The median increase is about $0.3$ kpc in the two quenched populations.
%Instead, for the star-forming sub-samples, we find a median variation of about $0.05$ kpc, but $\sim50\%$ of \SFPM{} galaxies experience a reduction of \rgal. 
%However, for a typical star-forming galaxy with \rgal{} $=10$ kpc and a gas mass surface density of $3$ \msol{} kpc$^{-2}$ \citep{Kennicutt1998}, a reduction of $0.4$kpc could account for an amount of lost gas of only $\sim10^8$\msol{}, well below the values we measure. 
Thus, we can rule out the conversion into stars as a cause of the observed reduction of gas in quenched galaxies, which instead can be only explained by gas ejected from the galaxy beyond \rgal{}.
%The high values of gas loss can be explained by high rates of gas loss, due to a feedback process able to eject the gas out of star-forming regions. 
We also find that $70.8\pm8.4\%$ of quenched post-merger galaxies remain quenched for the rest of the simulation (i.e. up to z$=0$),  $22.8\pm4\%$ of them exhibit only sporadic episodes of low star formation rates and only around $6.4\pm2\%$ of the quenched post-merger galaxies rejuvenate. These numbers are in accordance with the evolution of the quenched control population, with $72.4\pm4\%$ permanently quenched, $15.8\pm1.5\%$ that show some episodes of low star formation, and $11.8\pm1.3\%$ which return to the star-forming main sequence. The aforementioned results suggest that the the mechanism responsible for the gas removal must also be responsible for preventive feedback, which would explain the lack of rejuvenation.

%&&&

We showed earlier in this section that the star-forming post-merger and control populations experience less gas mass loss than quenched post-merger and control galaxies (see \autoref{fig:gasloss}), therefore, this would require converting $\sim1-2\times10^9$\msol{} of gas into stars in about $500$ Myr. 
\autoref{fig:RvsM} reveals that $\sim49.5\%$ of star-forming post-mergers and $\sim60\%$ of star-forming control galaxies have increased their stellar mass of at least $1\times10^9$\msol{} over $500$ Myr, an amount consistent with the measured gas loss.

%\autoref{fig:RvsM} shows that \SFPM{} and \SFCTRL{} populations have slightly higher stellar mass growth than the quenched populations, with a median of $\sim10^{9}$\msol{} in new stars formed (or accreted) within $500$Myr. 
%Since these star-forming galaxies experience less gas mass loss than \QPM{} and \QCTRL{} galaxies (see \autoref{fig:gasloss}), this would require converting $\sim3\times10^9$\msol{} of gas into stars in about $500$ Myr

In this section, we showed that both quenched post-merger and control galaxies show high rates of gas loss and a high fraction of permanent quenching. 
This common behaviour supports a scenario where the vast majority of both quenched post-mergers and quenched control galaxies are quenched because of a common feedback process that ejects the gas out of galaxies and prevents its further accretion by either supplying kinetic energy to the remaining medium or by increasing the entropy of the ejected gas and prolonging its cooling time \citep[see][]{Zinger2020}.
In the next section, we focus on the quenching mechanism in the quenched post-merger population and we compare the outcome with that of the quenched control galaxies, in order to understand the origin of the excess of quenched post-mergers in the early times after the coalescence phase.

%\begin{figure}
%\centering
%	\includegraphics[width=\columnwidth]{BHfeedbAndBindEn_v100.pdf}
%	\caption{(a) The cumulative energy injected in the medium as kinetic feedback from the SMBH (from the origin of the galaxy to $\sim 450$ Myr from \tpm{}) as a function of the SMBH mass. (b) the ratio between the cumulative energy in (a) and the total binding energy of the gas particles, as a function of  the SMBH mass. The post-mergers are represented by filled cirlces (red symbols for post-mergers that have quenched within $0.5$ Gyr from the merging, while blue symbols for post-mergers descendants that remains star-forming). Instead, the control galaxies are represented by open squares (red symbols for control galaxies that have quenched within $0.5$ Gyr from \tpm{}, while blue symbols for control galaxies descendants that remains star-forming).}
%    \label{fig:example_figure}
%\end{figure}

%%%%%%%%%%%%%%%%%%%%%%%%%%%%%%%%%
\subsection{The impact of AGN feedback}
\label{sec:AGN}
In the previous section, we showed that gas loss is the cause of quenching in post-merger galaxies, as well as controls, which suggests that ejective quenching mechanisms are responsible for the gas loss in both samples.
In this section, we explore the possibility that different mechanisms are responsible for the quenching in the post-mergers and control quenched galaxies.
In Section~\ref{sec:tng} we briefly described the AGN feedback implemented in TNG model.
To recap, the TNG AGN model feedback employs either pure kinetic feedback at low accretion rates, or thermal feedback at high accretion rates.
This scheme is in accordance with the two modes of activity in observed AGNs (\citealp[e.g.,][]{Crenshaw2010, VillarMartin2011, Woo2016}; for the high accretion mode, and \citealp{Fabian2012, McNamara2007}, for the low accretion mode), with improved agreement with the observational results regarding co-evolution of galaxies and black-holes \citep{Weinberger2017, weinberger2018}.
At high accretion rates, the TNG model
injects pure thermal energy into the gas surrounding the black hole. However, \citet{Weinberger2017} show that such energy does not efficiently couple with
the gas, resulting in
almost unaltered thermodynamics of the TNG gas cells, with no or insufficient impact on the cooling/heating functions and, therefore, on the star formation in TNG galaxies \citep{Weinberger2017, Terrazas2020}.
To simulate the feedback at low accretion rates, instead, TNG uses a kinetic wind model.  
The energy accumulates proportionally to the accretion rate until a threshold amount is reached.
Then, the kinetic energy is
released impulsively into the gas surrounding the black hole in a random direction
\citep{Weinberger2017}.

Studies of quenching in IllustrisTNG \citep{Weinberger2017, Nelson2018, Terrazas2020} reveal that only the kinetic mode of AGN feedback offers the necessary conditions to suppress the star formation in TNG galaxies with M$_\ast \geq 10^{10}$M$_\odot$, by pushing gas away from the
galaxy.
Moreover, \citet{Terrazas2020} demonstrate that the ejective feedback in TNG becomes effective at
quenching star formation once the cumulative kinetic energy
 overcomes the total gravitational binding energy of the
gas in a galaxy.
They also show that the kinetic feedback process dominates in galaxies whose M$_\text{BH}$ exceeds $10^{8.2}$M$_\odot$, the black hole mass threshold above which more than $90\%$ of the TNG galaxies are quenched
\citep[see also ][]{Zinger2020}.
This M$_\text{BH}$ threshold
for quiescence arises from the TNG model parameters chosen to reproduce observational properties
of the galaxy population at the present time \citep{Pillepich2018a}.

\begin{figure}
\centering
\includegraphics[width=\columnwidth]{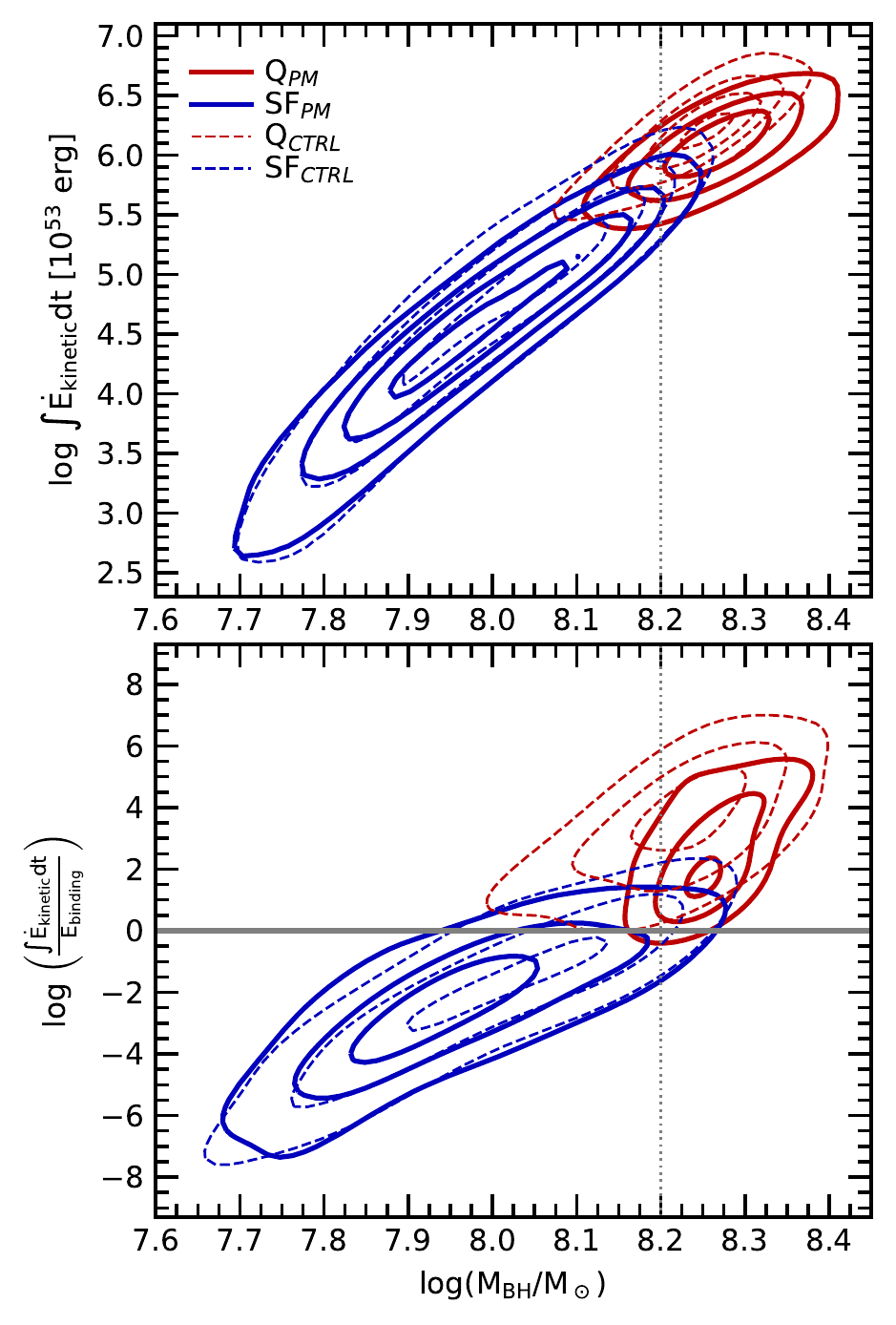}
	\caption{(Top) The cumulative kinetic energy injected in the gas surrounding the central black hole as a function of black hole mass. (Bottom) The ratio between the cumulative kinetic energy and the total gravitational binding energy of the gas, as a function of black hole mass. The post-merger galaxies are represented by solid contours: red for quenched post-mergers (\QPM{} galaxies), and blue for star-forming post-mergers (\SFPM{} galaxies). Control galaxies are represented by dashed contours: red for quenched control galaxies (\QCTRL{} galaxies), and blue for star-forming control galaxies (\SFCTRL{}). 
	 The vertical dotted grey line at M$_\text{BH}=10^{8.2} \text{M}_\odot$ represents the mass threshold above which more than $90\%$ of TNG galaxies are quenched \citep[e.g.,][]{Zinger2020}.
	The figure indicates that a necessary condition for star formation quenching in TNG galaxies is that the cumulative kinetic energy from the AGN feedback must overcome the total gravitational binding energy that keeps the gas bound to the galaxy.
	}
    \label{fig:agn_relations}
\end{figure}

In the top panel of \autoref{fig:agn_relations} we investigate the role of the AGN on quenching in our post-merger sample by showing the correlation between the cumulative kinetic energy released into the gas and the central black hole mass. 
Most of the quenched post-mergers sample galaxies occupy the high \bhm{} regime \bhm$\geq10^{8.2}$\msol{} and also have the largest cumulative kinetic energies. 
The star-forming post-merger galaxies instead preferentially have \bhm$<10^{8.2}$\msol{} and exhibit a wide range of cumulative kinetic energies. 
The behaviour of the TNG controls is broadly consistent with that of the quenched and star-forming post-merger populations, although
%, with the only difference that there is a larger number of \QCTRL{} galaxies at \bhm$<10^{8.2}$\msol{}. 
quenched post-merger galaxies have slightly  larger \bhm{} than quenched control galaxies, with $\sim80\%$ of quenched post-mergers exceeding the threshold of \bhm$=10^{8.2}$\msol{}, whereas only $\sim60\%$ of the quenched control galaxies have larger \bhm{} than the threshold value. 
Therefore neither the black hole mass threshold nor a large amount of cumulative kinetic feedback are sufficient conditions for quenching TNG galaxies.

\citet{Terrazas2020} show that the additional key ingredient necessary to understand star formation quenching in TNG galaxies is the proportion between the total amount of energy released into the gas via AGN kinetic feedback and the total gravitational energy felt by the gas cells. 
Indeed, \citet{Terrazas2020} showed that TNG galaxies whose kinetic energy overcomes the binding energy are typically quenched because the feedback has enough energy to push the gas out from the galaxies.

In the bottom panel of \autoref{fig:agn_relations}, we show the ratio between the total amount of energy released into the gas via AGN kinetic feedback and the total gravitational energy of the gas cells ($\tfrac{\int \dot{\text{E}}_\text{kinetic} \text{dt}}{\text{E}_\text{binding}}$), as a function of black hole mass. 
We find that the TNG post-mergers and controls follow the same trend, with a sharp transition when the kinetic energy overtakes the binding energy of the gas. 
%\begin{figure}
%\centering
%	\includegraphics[width=0.9\columnwidth]{BHfeedbAndBindEn_4plots_v110.pdf}
%	\caption{Cumulative distributions of four useful TNG quantities we use for the AGN feedback analysis: (a) the ratio between the total AGN kinetic energy feedback and the total binding energy of the gas particles; (b) the black hole mass; (c) total AGN kinetic energy feedback, and (d) total binding energy of the gas particles.  The solid curves represent the distributions of \SFPM{} (color blue) and \QPM{} (red), whereas the dashed ones represent the distributions of \SFCTRL{} (color blue) and \QCTRL{} (red). In all the panels, the horizontal continue grey line represents the median of the distribution, whereas the horizontal dashed grey lines are the $16^\text{th}-84^\text{th}$ percentiles enclosing $68\%$ of the distribution. The vertical dotted grey line in (a) highlights the position where the two energies are balanced, whereas the one in (b) represents a threshold black hole mass above which $90\%$ of TNG galaxies are quenched.}
%    \label{fig:cumul_energies}
%\end{figure}
%\autoref{fig:cumul_energies}(a) shows the energy ratio cumulative curves of the energetic ratio. 
%\textcolor{purple}{\sout{We find that  in $\sim82\%$ of \QPM{} and $\sim81\%$ of \QCTRL{}} \sout{galaxies the total kinetic energy is more than $10$} \sout{times higher than the binding energy, in contrast to the only $\sim4\%$ of \SFPM{}  and $\sim7\%$ of \SFCTRL{} galaxies.}}
In $\sim99\%$ of quenched post-mergers and in all quenched control galaxies the total kinetic energy is higher than the binding energy, in contrast to only $\sim12.7\%$ of star-forming post-mergers and $\sim22.7\%$ of star-forming control galaxies. 
Therefore, the distinction between quenched and still star-forming galaxies is cleaner when binding energy is considered, as already reported by \citet{Terrazas2020}. 
However, closer analysis of the bottom panel of \autoref{fig:agn_relations} shows that there are some differences between quenched post-mergers and quenched control galaxies.  
At fixed M$_\text{BH}$, quenched post-merger galaxies have, on average, lower $\tfrac{\int \dot{\text{E}}_\text{kinetic} \text{dt}}{\text{E}_\text{binding}}$ than quenched control galaxies, thus suggesting they have higher binding energy (because quenched post-mergers and quenched control galaxies have similar cumulative kinetic energy, see the top panel of \autoref{fig:agn_relations}).  
%\textcolor{orange}{\sout{Such difference in the average $\tfrac{\int \dot{\text{E}}_\text{kinetic} \text{dt}}{\text{E}_\text{binding}}$ of the two quenched populations}\sout{ might be related to the effect of merging galaxies. 
%Mergers could help the AGN kinetic feedback to quench star formation despite a strong gravitational} \sout{ potential. In other words, once the AGN kinetic feedback is operative, it is easier to quench a post-merger than a non-post-merger galaxy. }
%\sout{Such behaviour could also explain the excess of quenched post-mergers we found in Section~\ref{sec:ratio}. }}
Another difference between quenched post-merger and control populations is that there is a tail of quenched control galaxies with M$_\text{BH}<10^{8.2}$ M$_\odot$. 
We recall that the definition of quenched control galaxies includes all the quenched controls, hence the quenched control population represents the behaviour of the whole TNG quenched population.  
Therefore, though rare, there are quenched galaxies with black hole masses below the typical TNG mass threshold.

%%It is also worthwhile investigating the distribution of the two energy type independently.
%%\autoref{fig:cumul_energies}(c) and (d) show the total kinetic energy and binding energy cumulative curves, respectively. 
%The energy balance shown in \autoref{fig:agn_relations} uses the galaxy binding energy as a reference.  Given the morphological transformation of the post-merger sample, the rearrangement of mass could potentially make it harder for the galaxy to eject gas.  Indeed, we find that the distribution of the binding energy of \QPM{}  galaxies is shifted to higher values of binding energy compared to the distribution of \QCTRL{} galaxies. 
%Therefore, AGN feedback in \QPM{} galaxies needs to inject a larger amount of kinetic energy to overcome the gravitational binding energy of the gas. 
%However, despite the discrepancy between the two populations, the analysis of the distributions of the kinetic energy reveals that the AGN activity in \QPM{} galaxies is strong enough to overtake the binding energy.
%In fact, about $90\%$ of \QPM{} and $79\%$ of \QCTRL{} galaxies exceed cumulative energy of $10^{58.5}$ erg injected into the medium surrounding the central black hole, compared to about $7\%$ and $12\%$ of \SFPM{} and \SFCTRL{} galaxies, respectively.

The results in \autoref{fig:agn_relations} represent the instantaneous situation at the moment of  quenching in the quenched post-merger population. 
It is instructive to observe how the two energy types evolve with time. 
\begin{figure}
\centering
	\includegraphics[width=\columnwidth]{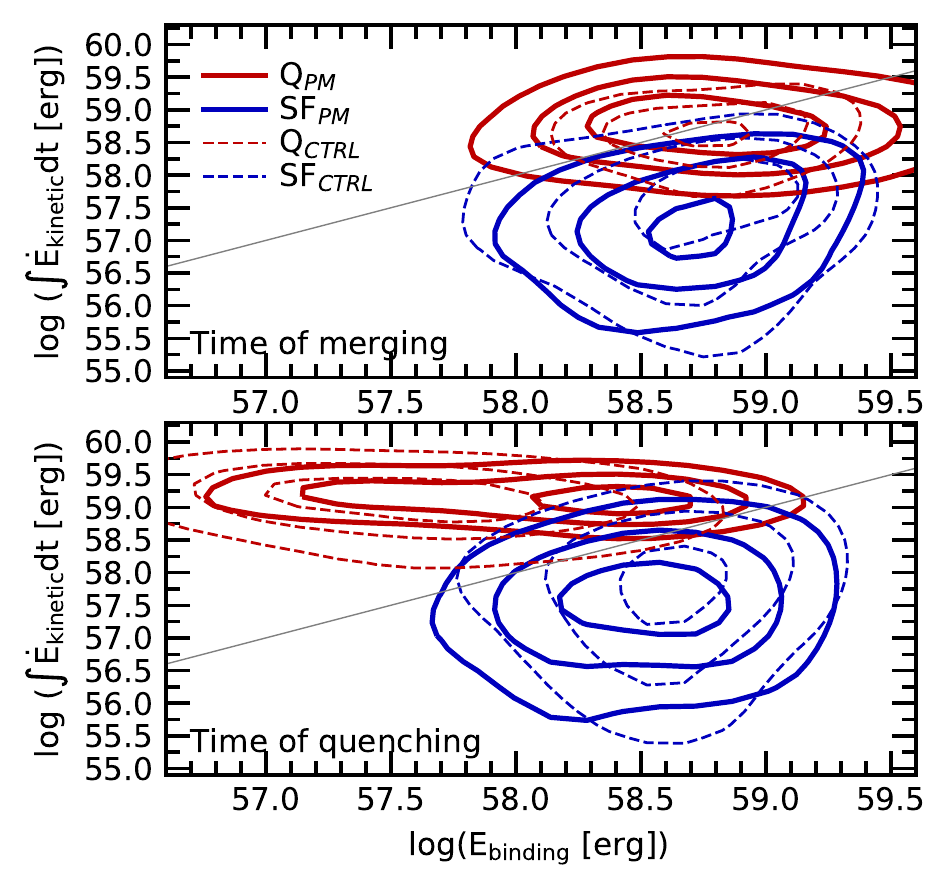}
	\caption{(Top) The relation between the cumulative energy injected in the medium from the AGN activity as kinetic feedback  (from the time when the black hole is seeded in the centre of the galaxy to \tpm$=0$) and the total binding energy of the gas particles at \tpm$=0$.  (Bottom) The same relation but at the time of quenching.	
	The post-merger galaxies are represented by solid contours: red for quenched post-mergers (\QPM{} galaxies), and blue for star-forming post-mergers (\SFPM{} galaxies). Control galaxies are represented by dashed contours: red for quenched control galaxies (\QCTRL{} galaxies), and blue for star-forming control galaxies (\SFCTRL{}). 
	The grey line represents the bisector at which the cumulative binding energy equals the total binding energy. 
	The figure shows that  in \QPM{} and \QCTRL{} galaxies the binding energy decreases
steadily, whilst the total kinetic feedback only slightly increases.
	}
    \label{fig:energyevol}
\end{figure}
\autoref{fig:energyevol} shows the relation between the cumulative kinetic energy and the binding energy at t$_\text{PM}=0$ (top panel) and at the time of the quenching (bottom panel): red for quenched post-mergers (\QPM{} galaxies), and blue for star-forming post-mergers (\SFPM{} galaxies). Control galaxies are represented by dashed contours: red for quenched control galaxies (\QCTRL{} galaxies), and blue for star-forming control galaxies (\SFCTRL{}). 

At t$_\text{PM}=0$ (\autoref{fig:energyevol}, top panel), we find the quenched post-mergers and quenched control galaxies show, on average, similar values of binding and kinetic energies.  Conversely, for the star-forming post-mergers and star-forming control galaxies the binding energy is larger than the kinetic energy, hence the gas is retained for ongoing star formation.
Instead, at the time of quenching (\autoref{fig:energyevol}, lower panel), we find that both quenched post-mergers and quenched control galaxies have reduced binding energy, by typically $-0.5$ dex for quenched post-mergers and $-1.1$ dex for quenched control galaxies, while increasing their kinetic energy by $\sim0.3$ dex.

Following the evolution of the two energy types (as in \autoref{fig:energyevol}) between t$_\text{PM}=0$ and the time of quenching, we find that the binding energy decreases steadily, while the total kinetic feedback only slightly increases.
It is worth noting that there is a significant fraction of star-forming galaxies that exhibit an energetic ratio $\tfrac{\int \dot{\text{E}}_\text{kinetic} \text{dt}}{\text{E}_\text{binding}} > 1$, as the quenched populations (see bottom panels of \autoref{fig:agn_relations} and \autoref{fig:energyevol}).
This means that the energy balance between kinetic and binding energies is not by itself a sufficient condition for quenching in TNG galaxies.
By definition (see equation~\ref{eq:bind}), the binding energy is proportional to the amount of gas in the reservoir of the galaxies. 
Consequently, it is expected that the binding energy decreases subject to the high gas loss we find in the quenched galaxy population in TNG (see Section~\ref{sec:gasevol}). 
Moreover, the binding energy of the gas depends also on the strength of the gravitational potential, hence on the global mass distribution. 
Therefore, the binding energy, by definition, reflects the gas fraction in galaxies.
%\textcolor{orange}{\sout{In fact, once the AGN feedback becomes effective in TNG galaxies,}\sout{ it has sufficient energy to win over the binding energy of some gas cells within the galaxy, and the total} \sout{ binding energy decreases (see equation~\ref{eq:bind}).}\sout{ Then, progressively less energy would be required to overtake the binding energy of the remaining gas.}} 
%Moreover, by reducing the amount of gas in the reservoir, the resulting accretion onto the black hole decreases, resulting in the feedback being in the kinetic rather than the radiative mode. 
\begin{figure}
\centering
	\includegraphics[width=\columnwidth]{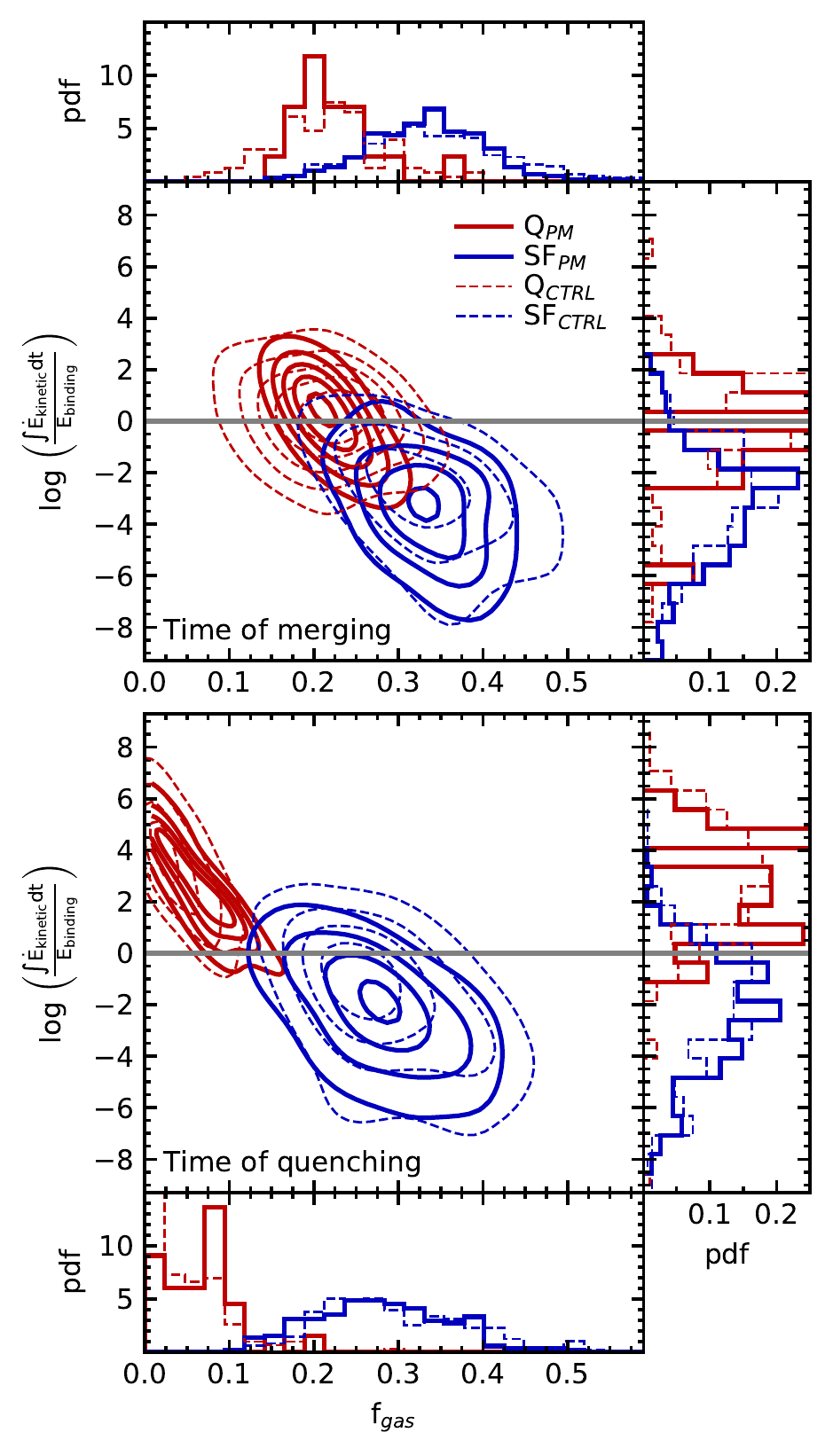}
	\caption{(Top) The ratio between the cumulative kinetic energy (released from the time when the black hole is seeded in the centre of the galaxy to the time of coalescence \tpm$=0$) and the total gravitational binding energy of the gas at  \tpm$=0$, as a function of the gas fraction at \tpm$=0$.
 (Bottom) The same relation but at the time of quenching (in the cases of the \QPM{} and \QCTRL{} populations), or \tpm$=500$Myr (in the cases of the \SFPM{} and \SFCTRL{} populations).
 	The post-merger galaxies are represented by solid contours: red for quenched post-mergers (\QPM{} galaxies), and blue for star-forming post-mergers (\SFPM{} galaxies). Control galaxies are represented by dashed contours: red for quenched control galaxies (\QCTRL{} galaxies), and blue for star-forming control galaxies (\SFCTRL{}). 
 	The figure demonstrates that TNG post-merger galaxies (but also non-post-merger galaxies)  quench when their gas fraction is below roughly $0.1$. 
	}
    \label{fig:energyVsGasFrac}
\end{figure}

\autoref{fig:energyVsGasFrac} shows the relation between the energetic ratio $\tfrac{\int \dot{\text{E}}_\text{kinetic} \text{dt}}{\text{E}_\text{binding}}$ and the gas fraction (i.e. the fraction of the whole gas mass over the baryonic mass, f$_\text{gas}$) at the time  t$_\text{PM}=0$ of merging (top panel) and at the time of quenching (bottom panel). 
At the time of merging (\autoref{fig:energyVsGasFrac}, top panel), when all our galaxies are still star-forming, the gas fractions of the populations that will eventually quench (i.e. \QPM{} and \QCTRL{}) are lower than those that remain star-forming (i.e. \SFPM{} and \SFCTRL{}).
Quantitatively, $\sim50\%$ of \QPM{} and \QCTRL{} galaxies  have f$_\text{gas}\lesssim0.2$, whilst \SFPM{} and \SFCTRL{} galaxies have instead gas fraction in the range $0.2\lesssim\text{f}_\text{gas}\lesssim0.5$.
This fact suggests that there is a pre-disposition towards quenching if the gas fraction is low to start with.

%, and almost all of them have $\tfrac{\int \dot{\text{E}}_\text{kinetic} \text{dt}}{\text{E}_\text{binding}}<1$. 
%We recall that the initial gas mass distributions of \QPM{} and \QCTRL{} populations do not differ from those of the \SFPM{} and \SFCTRL{} populations (see \autoref{fig:initial_gas}), therefore, a low level of gas fraction indicates that the \QPM{} and \QCTRL{} galaxies are more massive in stellar mass than the \SFPM{} and \SFCTRL{} galaxies.
%Moreover, we find that most of these \QPM{} and \QCTRL{} galaxies with low gas fraction have $\tfrac{\int \dot{\text{E}}_\text{kinetic} \text{dt}}{\text{E}_\text{binding}}>1$, thus suggesting that in these galaxies the AGN kinetic feedback should be already in place at t$_\text{PM}=0$ (i.e. when they are still star-forming galaxies). 

At the time of quenching (\autoref{fig:energyVsGasFrac}, lower panel), the separation between the quenched and star-forming populations is seen as a very sharp distinction in gas fraction. 
%We find that in all \QPM{} and \QCTRL{} galaxies, gas fraction  is f$_\text{gas}<0.2$, while it remains almost unchanged in the \SFPM{} and \SFCTRL{} populations.
We find a gas fraction threshold at roughly f$_\text{gas}<0.1$, below which all the TNG galaxies are quenched, and above which more than $98\%$ of galaxies are still star-forming. 
Therefore, the gas fraction is a better discriminator than the energy ratio between TNG quenched and star-forming galaxies.

In summary of this section, we find that quenching is rare amongst post-mergers, although still more frequent than in the control sample.  
The quenching process in both the post-mergers and controls is linked to loss of gas that is triggered by AGN feedback.
%, whose impact is a balance between the binding and kinetic energies. 
The star formation quenching dominates when the gas fraction is below f$_\text{gas}\lesssim0.1$.

%%%%%%%%%%%%%%%%%%%%%%%%%%%%%%%%
\section{Impact of \bhm{} matching scheme}
\label{sec:bhmismatch}
%In Section~\ref{sec:ratio} we find quenched galaxies to be only a small fraction amongst the descendants of star-forming post-mergers in TNG. 
%In Section~\ref{sec:AGN}, we also show that the mechanism acting to interrupt the star formation is strictly connected to the AGN activity, with no substantial differences between post-mergers and control galaxies.
A key part of our experimental set up is the construction of the control sample.  Our fiducial scheme includes matching in stellar mass, SFR, environment and black hole mass.  
In particular, in order to limit any bias related to AGN in post-mergers and their controls, our matching criteria includes a maximum tolerance of $0.05$ dex (i.e. $\sim12\%$) on black hole mass of control galaxies (see Section~\ref{sec:method}). 
Nonetheless, we find an excess of quenched galaxies in post-mergers compared to control galaxies that have not experienced any merger in the last $2$ Gyr. 
Therefore, it is worth investigating whether the excess is real or if it depends on the chosen \bhm{} tolerance, or in other words, to what extent the \bhm{} matching could bias our results.
%This test consists of measuring the excess of post-mergers while varying the matching tolerance on \bhm{}. 

In order to test the impact of our matching scheme, we re-measure the excess of post-mergers for different M$_\text{BH}$ matching tolerances.
We first remove any restrictions on \bhm{} in our search for control galaxies, then we selected three further control samples with more stringent tolerance in the M$_\text{BH}$ difference than in the fiducial case, with $0.02$ dex (i.e. a maximum mismatch around $5\%$), $0.015$ dex (i.e. a maximum mismatch of roughly $3.5\%$), and $0.01$ dex (i.e. a maximum mismatch of about $2.3\%$). 
\begin{figure}
\setlength{\lineskip}{-16.5pt}
\centering	\includegraphics[width=0.85\columnwidth]{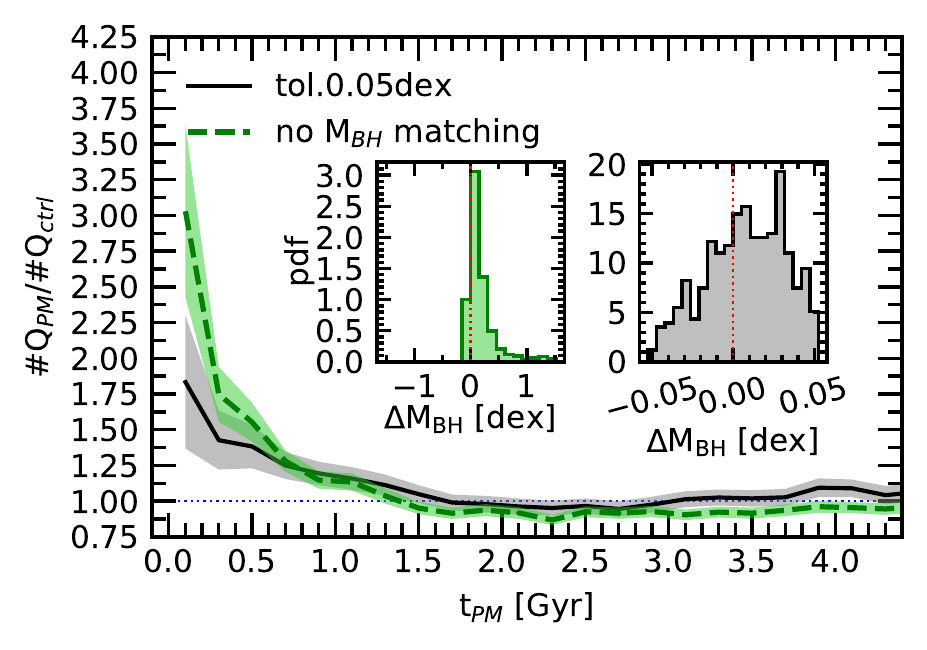}	\includegraphics[width=0.85\columnwidth]{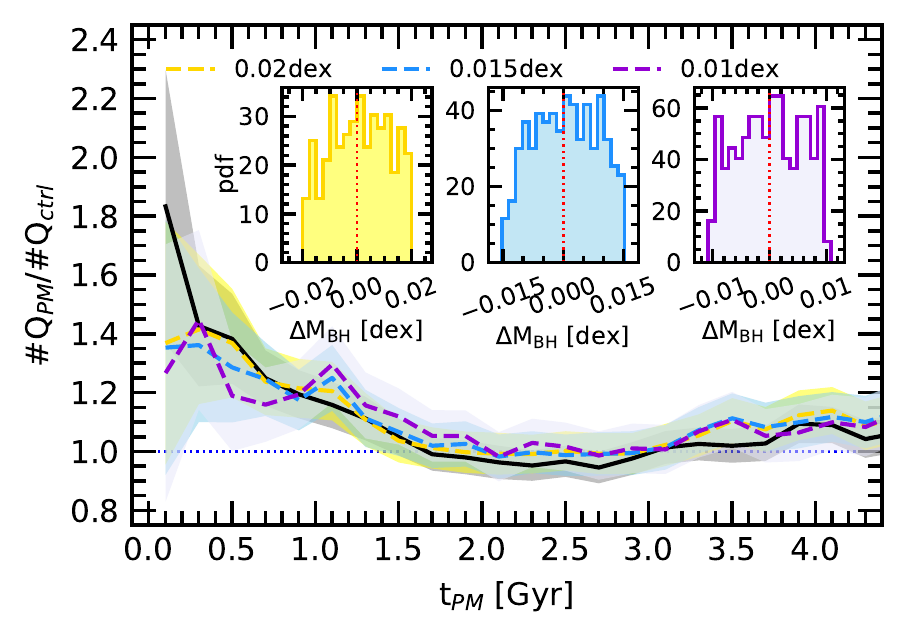}
	\caption{The ratios between the number of quenched post-mergers (\QPM{}) and quenched controls (\QCTRL{}) as a function of time elapsed since coalescence t$_\text{PM}$. The black curve and grey contours in both panels represent the reference result we show in \autoref{fig:ratio_evolution}, corresponding to a maximum tolerance of $0.05$ dex (i.e. $\sim12\%$) in the match of black hole masses of control galaxies. The green curve in the top panel represents the ratio when none black hole matching criterion is applied.  The coloured curves in the bottom represent the ratio relative to three tighter tolerance limits than that we choose in our analysis: yellow for a maximum mismatch of $0.02$ dex (i.e. within $5\%$), blue for $0.015$ dex (i.e. a maximum mismatch of $3.5\%$) and purple for $0.01$ dex (i.e. within $2.3\%$).
	The insets in each panel show the distributions of $\Delta$M$_\text{BH}$ (i.e. M$_\text{BH}$ of PMs - M$_\text{BH}$ of CTRLs).
	The figure demonstrates that TNG post-mergers have a higher chance to quench the star formation in the early stages after coalescence compared to non-post-merger galaxies with similar characteristics and environment. 
The excess of quenched post-mergers cannot be simply ascribed to a mismatch in the black hole mass of the two populations.}
    \label{fig:excessComp}
\end{figure}
The top panel of \autoref{fig:excessComp} shows the ratio between the number of quenched post-mergers and quenched control galaxies matched without any restrictions on \bhm{}, while the bottom panel of \autoref{fig:excessComp} shows the same ratio for each of the three cases with more stringent tolerance on \bhm{} match.
As a reference, in both panels we also display the ratio obtained with our fiducial tolerance, as in \autoref{fig:ratio_evolution}.
We find that the different matchings show qualitatively consistent results, with an excess of quenched post-mergers immediately after coalescence followed by a steadily decreasing ratio with the time passed after the merger. 
Therefore, our qualitative conclusion, that mergers lead to an excess of quenched galaxies, is robust against the choice of the tolerance criterion on \bhm{}.
However, changing the matching tolerance has a quantitative effect on the quenched fraction in the two populations. 
In the case where we do not match in black hole mass, we find a larger excess immediately after the coalescence phase (i.e. $\leq160$ Myr after the merger) compared to Section~\ref{sec:ratio}. 
The excess of quenched galaxies when no \bhm{} matching is used is $3\pm0.6$ times, at a significance level of $3.3\sigma$ (where $\sigma$ represents the Poissonian error of the ratio). 
In the three cases with lower tolerances, the excess is $1.4\pm0.6$ times the number of \QCTRL{} galaxies in all the three cases, only slightly smaller than the excess referred to the fiducial case, and the excess is confirmed at a significance level of $\sim1\sigma$. 
We point out that by requiring more stringent constraints on the black hole mass of the control galaxies we reduce the chance of finding a control galaxy for each post-merger, with the drawback of worsening the statistics of our result. 
Indeed, only $3\%$ of TNG post-mergers have a control galaxy that can be matched in M$_\text{BH}$ with a tolerance of $0.01$ dex. 
However, we find consistent trends for all tolerances considered, thus suggesting that the excess in the early phase after the merger would be present even with more extreme constraints on \bhm{} of matched control galaxies. 

We next investigate the origin of the larger excess of post-mergers in the case without matching on black hole mass. Since AGN feedback is the driver of quenching in TNG, we expect that the higher fraction of quenched post-mergers compared with control galaxies should be related to an excess of post-mergers with M$_\text{BH}\geq10^{8.2}$\msol{}, that is the black hole mass threshold above which more than $90\%$ of the TNG galaxies are quenched (see Section~\ref{sec:AGN}) and that can sustain efficient kinetic feedback.
The inset panels in \autoref{fig:excessComp} show the distributions of the difference between the M$_\text{BH}$ of TNG post-mergers that have M$_\text{BH}\geq10^{8.2}$\msol{} and their controls ($\Delta$M$_\text{BH} = $ M$_\text{BH}$ of PMs - M$_\text{BH}$ of CTRLs). 
When we do not match in black hole mass, we find a skewed $\Delta$M$_\text{BH}$ distribution, with almost all the post-mergers with M$_\text{BH}\geq10^{8.2}$\msol{} matched to control galaxies with lower black hole mass. 
The median of the distribution is $\Delta$M$_\text{BH} = 0.11$ dex, but the high-end tail of the distribution show that $\sim5\%$ of post-mergers with massive black holes have $\Delta$M$_\text{BH} > 1$ dex, thus matched to controls whose black hole activity cannot sustain efficient AGN kinetic feedback. 
This test suggests that post-mergers with high-mass black holes have less chance to be matched to a control with comparable black hole mass. 
In other words, post-mergers could have a tendency to have higher black hole mass at fixed stellar mass than the non-post-merger population, that is what we might expect if mergers lead to enhanced black-hole accretion (Byrne-Mamahit et al., in preparation).
%The over-abundance of high-massive black holes and the link to the TNG AGN feedback model explain the larger excess of \QPM{} in the early post-merger stages, and it could be justified by the merger of the progenitors' black holes. 
The inset panels in \autoref{fig:excessComp} also show the $\Delta$M$_\text{BH}$ distribution related to the fiducial case we use in the rest of this paper (i.e. \bhm{} tolerance of $0.05$ dex, or $\sim12\%$) and the $\Delta$M$_\text{BH}$ distributions of the three cases with progressively reduced tolerance on \bhm{} of control galaxies.
The $\Delta$M$_\text{BH}$ distribution of the reference case has a median at $\sim0.01$ dex. 
Therefore, also in the reference case there are slightly more post-mergers matched to controls with less massive black holes, however, with a maximum difference of only $0.05$ dex in \bhm{}, the AGN model guarantees control galaxies with similar AGN feedback. 
The $\Delta$M$_\text{BH}$ distributions of the three cases with progressively reduced tolerance on \bhm{} of control galaxies are symmetric, further limiting any mismatch in \bhm{}. 

%We tress that the excess of \QPM{} is still present, though reduced, in the reference case, where we force the control galaxies to have black holes masses within $0.05$ dex, or $\sim12\%$, and also in the cases with tighter constrain , despite the mismatch distribution is tighter and symmetric (see the inset panels in \autoref{fig:excessComp}).
%However, a tolerance of $\sim12\%$ in M$_\text{BH}$ is large enough to allow $35$ \QPM{} galaxies to have a black hole mass 

To summarise, the results presented in this section suggest that TNG post-mergers have a higher chance to quench the star formation in the early stages after coalescence compared to non-post-merger galaxies with similar characteristics and environment. 
The excess of quenched post-mergers cannot be simply ascribed to a mismatch in the black hole mass of the two populations. 
Indeed, the excess persists when we remove any matching on the black hole mass. 
Therefore, even if the quenching in TNG post-mergers is strictly connected to AGN activity (see Section~\ref{sec:AGN}), the dynamics of galactic mergers could contribute to halting star formation in post-mergers.
%For example, \citet{Pontzen2017} applied a genetic modification approach \citep{Roth2016} to generate sets of controlled numerical realisations in a fully cosmological context of a halo of $10^{12}$\msol{}, by altering its accretion history. 
%They find that in major mergers (a mass ratio of 2:3) AGN feedback alone is not sufficient to permanently quench star formation, but it acts in synergy with the kinetic effects of the merger. 
%The interaction disrupts the gaseous disk of the galaxy, resulting in a turbulent medium able to remove angular momentum from inflowing material; then, the inflowing material can easily reach the galactic centre to feed the black hole and the subsequent AGN activity contributes significantly to removing the remaining gas from the galaxy \citep[see also][]{Chadayammuri2020}.
%
%More recently, \citet{Sanchez2020} used a similar approach to analyse the impact of minor mergers on the star formation of simulated Milky Way analogues. They find that two small satellites interacting with the host can quench a Milky Way-like galaxy. The mechanism is similar to the one in \citet{Pontzen2017}, but in this case, it requires a tandem operation of a merger with the first satellite and subsequent close interaction with the second satellite to disrupt the gaseous disk and trigger intense AGN activity to halt the star formation.
%In a follow-up project, we will statistically analyse the repercussions of multiple interactions/mergers on quenching in large cosmological simulations.

%%%%%%%%%%%%%%%%%%%%%%%%%%%%%%%%%%%%%%%%%%%%%%%
\section{Resolution effects}
\label{sec:reseffects}
In the work presented thus far, we exclusively analyse the evolution of post-mergers from TNG300-1. 
TNG300-1 offers the most robust statistics (a larger number of post-mergers) while maintaining a reasonable spatial and mass resolution.
In this section, we perform a convergence test to investigate the robustness of our results against changes in the simulation's resolution (\citealp[see][]{Pillepich2018a} for details about the convergence of the IllustrisTNG physical model).
IllustrisTNG offers two other flagship simulations that use the same physical model but have different spatial and mass resolutions. 
The simulation TNG100-1 has $2\times2500^3$ resolution elements, and a dark matter particle mass resolution of $m_\text{dm} = 7.5\times10^6$\msol{} and a baryonic target mass $m_\text{bar} = 6\times10^6$\msol{}, respectively, which corresponds to approximately an order of magnitude better spatial and mass resolution than TNG300-1. TNG100-1 simulates a smaller volume of $110.7^3$ Mpc$^3$, that is around $1/20^\text{th}$ of the volume simulated in TNG300-1 (see Section~\ref{sec:tng}). The second simulation, TNG100-2, has the same volume as TNG100-1, but roughly the same mass and spatial resolutions as TNG300-1. 
Analysing the three simulations allows us to perform a test on the robustness of our results against distinct spatial and mass resolutions. 

We note that TNG100-1 includes a total of $1855$ post-mergers in the redshift range between $0\leq\text{z}\leq1$ , whereas TNG300-1 provides a sample of $25836$ post-mergers \citep[see][]{Hani2020}. 
By applying to TNG100-1 the same matching prescription we use for TNG300-1 (that consists of finding a control sample to the star-forming post-mergers with a match in the six parameters of redshift, stellar mass, SFR, N$_2$ and r$_1$ and \bhm{}), we would find control galaxies for only $64$ star-forming post-mergers (compared with $3472$ in TNG300-1).
We find that the requirement on \bhm{} is the one that most significantly reduces the yields of controls.
However, in Section~\ref{sec:bhmismatch} we showed that including a match in black-hole mass only impacts the excess of quenched post-mergers slightly, therefore, to increase the number of star-forming post-mergers in TNG100-1 and TNG100-2 and improve the statistics of the result, we compare the behaviour of the three simulations without matching the control samples in black hole mass, but only in the other five aforementioned parameters. 
This way, we obtain a sample of $252$ and $261$ star-forming post-mergers in TNG100-1 and TNG100-2, respectively.

\begin{figure}
	\includegraphics[width=\linewidth]{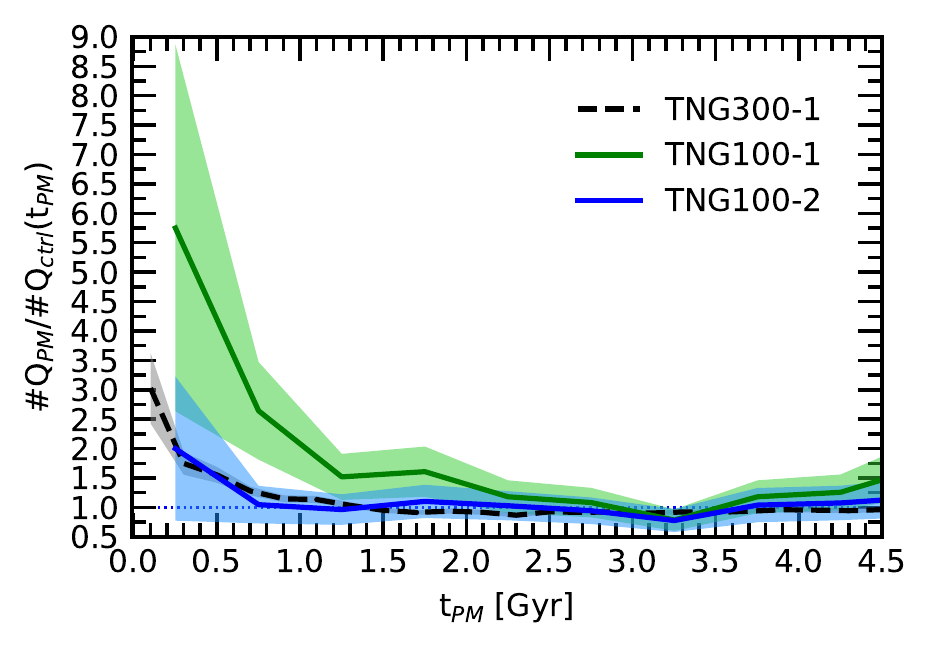}
	\caption{The ratio between the number of quenched post-mergers (\QPM{}) and quenched controls (\QCTRL{}) as a function of time elapsed since coalescence (t$_\text{PM}$) for three TNG simulations at different mass and spatial resolutions, and different volumes. The black curve and grey shading represent the result from TNG300-1 without any match in black-hole mass. The green solid curve with light green shading, and the blue solid curve with light blue shading represent the ratios obtained from TNG100-1 and TNG100-2, respectively. 
	To improve the statistics, the ratios from TNG100-1 and TNG100-2 are evaluated in bins of \tpm{}$500$ Myr, instead of \tpm{}$160$ Myr as is done for TNG300-1.
	The figure reveals that the resolution of the simulation significantly affects the result, whilst the volume of the distribution affects only the statistics of the result but not the trend.}
    \label{fig:excessResol}
\end{figure}
\autoref{fig:excessResol} shows the ratio between the number of quenched post-mergers (\QPM{}) and quenched control galaxies (\QCTRL{}) as a function of t$_\text{PM}$ for the three TNG simulations. TNG300-1 and TNG100-2, the two simulations at a similar numerical resolution, show a very similar result, thus
the reduced volume in TNG100-2 has little impact on the results other than poorer statistics (i.e., larger errors in the \QPM{}/\QCTRL{} ratio).
TNG100-1 shows qualitatively similar results to the lower resolution counterparts, with a grater excess (at a $1\sigma$ significance level) of quenched post-merger galaxies with respect to the quenched control sample.
The difference in \QPM{}/\QCTRL{} between the two resolution levels arises from a larger percentage of post-mergers that quench in the early phase after coalescence in the higher resolution simulation (TNG100-1). 
Around $250$ Myr after coalescence we find that  $\sim7\%$ of post-mergers are quenched in TNG100-1, compared to $\sim3\%$ in TNG300-1 and $\sim4\%$ in TNG100-2, while the quenched fraction in the control samples is comparable regardless of the simulation's resolution ($\sim1.5\%$).
 Understanding the dependence of \QPM{}/\QCTRL{} on the simulation's resolution is beyond the scope of this work. Nonetheless, qualitatively all the three simulations offer compatible results, suggesting that our findings are robust to changes in the simulations' mass and spatial resolutions. 

%%%%%%%%%%%%%%%%%%%%%%%%%%%%%%%%%%%%%%%%%%%%%%%
\section{Summary and conclusions}
\label{sec:conclusions}
In this work, we present an analysis of the incidence and the causes of quenching in IllustrisTNG post-merger galaxies. 
We quantify the impact of single mergers on quenching star formation within $500$ Myr after coalescence. We follow the evolution of the star formation rate in the post-merger descendants, and we compare the evolution to that of a control sample of non-post-merger galaxies matched in redshift, stellar mass, SFR, black-hole mass, environment, and isolation. 
%Both post-merger and control galaxies did not experience any merger $2$ Gyr before  and we interrupt following their evolution when the post-merger or its own control galaxy undergoes a new close encounter or a merging with a mass ratio larger than $0.1$. 
Our findings can be summarised as follows:
\begin{itemize}
\item \textbf{Quenching in TNG post-mergers:} quenching is rare among the descendants of star-forming post-merger galaxies.
Only around $5\%$ of TNG post-merger galaxies quench within $500$ Myr after merging (see \autoref{fig:percent_evolution}). 
%As a reference, in the control sample the $\sim2.4\%$ of galaxies (\QCTRL{}) quench in the same time interval. 

\item \textbf{The excess of quenched post-mergers:} although quenching in post-mergers is rare, quenching occurs in post-mergers at twice the rate of the controls (see \autoref{fig:ratio_evolution}), i.e. $2.4\%$ of the control sample is quenched within $500$ Myr.

%We did not find any difference between the quenching process in the \QPM{} and the \QCTRL{} galaxies.
%However, the few per cent of post-mergers which quench within $500$ Myr from the coalescence are around twice the expected number from non-post-merger control galaxies.
The excess of quenched galaxies within the post-merger sample dissipates with time post-merger.
After $\sim1.5$Gyr, the post-merger quenched population is statistically indistinguishable from the control's quenched population (see Section~\ref{sec:ratio}). 
%The excess is not permanent and it follows a trend with a maximum in the snapshot immediately after the coalescence, that is followed by a decrement towards \tpm$\sim1.5$Gyr from the merging. Then, the number of \QPM{} and \QCTRL{} galaxies becomes statistically indistinguishable for the rest of the simulation lifetime.

\item \textbf{The effect of AGN feedback:} the kinetic mode of AGN feedback is responsible for quenching post-mergers in the TNG model. 
The feedback acts in two ways: (1) the kinetic feedback injects momentum into the gas particles surrounding the central black hole, thus ejecting the gas from the galaxy (ejective feedback, see \autoref{fig:gasloss}), and (2) the kinetic feedback also prevents the gas from cooling to replenish the galactic reservoir for sustaining new episodes of star formation (preventive feedback, see Section~\ref{sec:gasevol}).
Quenching is most effective in galaxies with M$_\text{BH} \gtrsim10^{8.2}$\msol{}. At black hole masses higher than M$_\text{BH} \gtrsim10^{8.2}$\msol{} the total kinetic energy injected into the gas by the central black hole dominates over the gas's gravitational binding energy; as a result a notable fraction of the gas reservoir is removed from the galaxy (see \autoref{fig:agn_relations}, and \autoref{fig:energyevol}).

\item \textbf{Quenching and gas fraction:}
We find that: (1) there is a pre-disposition towards quenching if the gas fraction in a galaxy is low, and (2) the separation between the quenched and star-forming populations at the time of quenching is demarcated by a very sharp distinction in gas fraction, as quenched galaxies dominate at gas fractions below f$_\text{gas} \sim 0.1$ (see \autoref{fig:energyVsGasFrac}).
 %We find a gas fraction threshold at roughly f$_\text{gas} < 0.1$, below which all the TNG galaxies are quenched, and above which more than 98% of galaxies are still star-forming. Therefore, the gas fraction is a better discriminator than the energy ratio between TNG quenched and star-forming galaxies.
%The quenching starts when the AGN activity in galaxies with a M$_\text{BH} \gtrsim10^{8.2}$\msol{} induces an efficient AGN feedback able to remove a massive amount of gas particles from the galaxies. Then, the star formation halts when the total kinetic energy released by the central black hole overtakes the total gravitational binding energy of the residual gas particles in the galaxy.

\item \textbf{The effect of different \bhm{} matching schemes:} the excess of quenched post-mergers with respect to the number of quenched galaxies in the control population is stronger when we do not include a matching criterion for the black hole mass.
Nonetheless, though slightly weaker, the excess persists also when we match control galaxies with a maximum tolerance of $0.015$ dex in black hole mass (see \autoref{fig:excessComp}).

%\item \textbf{The excess of quenched post-mergers is temporary:} 
\end{itemize}

The picture that arises from our analysis is that mergers in TNG do not contribute significantly to the quenching of star-forming galaxies. 
The rarity of quenching in post-mergers is qualitatively in accordance with other cosmological simulations and observational results. For instance, 
\citet{RodriguezMontero2019} find that major mergers in the SIMBA simulation \citep{Dave2019} are not directly related to quenching, as the typical delay between the merger and subsequent quenching is larger than $1$ Gyr.
By analysing a sample of galaxies extracted from the SDSS DR7, \citet{Weigel2017} show that major mergers should not be the preferred path leading to permanent quenching. They found that major merger quenched galaxies account for a maximum of $5$\% of the quenched population at a given stellar mass, both at low- and intermediate-redshift. 
%The $5\%$ of post-mergers that quench immediately after the coalescence is, however, larger than the $3\%$ expected from the control sample,

Despite the small absolute fraction of promptly quenched galaxies, there is a notable excess of quenched post-mergers compared to the control population, thus suggesting that mergers could have a non-negligible contribution to the quenching of star formation. 
Mergers disturb the internal kinematics of the galaxy (i.e. dark matter, stars, gas). 
The induced turbulence in the ISM could dissipate angular momentum of the infalling gas, and we are investigating this process in TNG post-mergers in a follow-up project. 
Gas infalling with reduced angular momentum could reach the centres of the galaxies and feed the central black holes. 
%\textcolor{orange}{\sout{However, if the black hole is not massive enough (i.e. $\gtrsim10^{8.2}$\msol{}), kinetic feedback is not strong enough to prevail over the gravitational binding energy of the gas. }
%\sout{The binding energy of the gas is larger than the kinetic feedback in most of the TNG star-forming post-mergers.}}
The temporary excess of quenched post-mergers found in the early phase after coalescence completely vanishes after $\sim1.5$Gyr following the merger, suggesting that the mergers accelerate the quenching process in those post-mergers whose progenitors were close to sustaining effective AGN kinetic feedback.
% by increasing the black hole mass, both via a black hole - black hole mergers and accretion. 

Our results may be dependent on the AGN feedback model implemented in IllustrisTNG. 
However, the dynamics of galactic mergers could contribute to halting star formation in post-mergers for a variety of reasons.
For example, \citet{Pontzen2017} applied a genetic modification approach \citep{Roth2016} to generate sets of controlled numerical realisations in a fully cosmological context of a halo of $10^{12}$\msol{}, by altering its accretion history. 
They find that in major mergers (a mass ratio of 2:3) AGN feedback alone is not sufficient to permanently quench star formation, but it acts in synergy with the kinetic effects of the merger. 
The interaction disrupts the gaseous disk of the galaxy, resulting in a turbulent medium able to remove angular momentum from inflowing material; then, the inflowing material can easily reach the galactic centre to feed the black hole and the subsequent AGN activity contributes significantly to removing the remaining gas from the galaxy \citep[see also][]{Chadayammuri2020}.
More recently, \citet{Sanchez2020} used a similar approach to analyse the impact of minor mergers on the star formation of simulated Milky Way analogues. They find that two small satellites interacting with the host can quench a Milky Way-like galaxy. The mechanism is similar to the one in \citet{Pontzen2017}, but in this case, it requires a tandem operation of a merger with the first satellite and subsequent close interaction with the second satellite to disrupt the gaseous disk and trigger intense AGN activity to halt the star formation.
In a follow-up project, we will statistically analyse the repercussions of multiple interactions/mergers on quenching in large cosmological simulations.
In a follow-up project, in order to gain new insights on the impact of mergers on quenching, we will analyse other cosmological simulations that implement different models to regulate black hole accretion, such as the Eagle simulation \citep{Schaye2015} and the Illustris simulation \citep{Vogelsberger2014}. 
%Moreover, we want to extend the study by including observational pieces of evidence of the interplay between the kinematical effects of mergings and AGN activity. 

\section*{Acknowledgements}
The authors are grateful to Dan Walters and Brian A. Terrazas for useful discussions. SQ is also grateful to Lucia Pozzetti, Douglas Rennehan and Isabel Santos Santos for helpful discussion and suggestions. 
MHH acknowledges support from the William and Caroline Herschel Postdoctoral Fellowship fund, and the Vanier Canada Graduate Scholarship.
DRP and SLE gratefully acknowledge NSERC for
Discovery Grants which helped to fund this research.
The simulations of the IllustrisTNG project used in this work were undertaken with compute time awarded by the Gauss Centre for Supercomputing (GCS) under GCS Large-Scale Projects GCS-ILLU and GCS-DWAR on the GCS share of the supercomputer Hazel Hen at the High Performance Computing Center Stuttgart (HLRS), as well as on the machines of the Max Planck Computing and Data Facility (MPCDF) in Garching, Germany.
This research was enabled, in part, by the computing resources provided by WestGrid and Compute Canada.

\section*{DATA AVAILABILITY}
The data used in this work are publicly available at \url{https://www.tng-project.org}.



\bibliographystyle{mnras}
\bibliography{bibTNG_vAccepted} % if your bibtex file is called example.bib

\begin{thebibliography}{}
\makeatletter
\relax
\def\mn@urlcharsother{\let\do\@makeother \do\$\do\&\do\#\do\^\do\_\do\%\do\~}
\def\mn@doi{\begingroup\mn@urlcharsother \@ifnextchar [ {\mn@doi@}
  {\mn@doi@[]}}
\def\mn@doi@[#1]#2{\def\@tempa{#1}\ifx\@tempa\@empty \href
  {http://dx.doi.org/#2} {doi:#2}\else \href {http://dx.doi.org/#2} {#1}\fi
  \endgroup}
\def\mn@eprint#1#2{\mn@eprint@#1:#2::\@nil}
\def\mn@eprint@arXiv#1{\href {http://arxiv.org/abs/#1} {{\tt arXiv:#1}}}
\def\mn@eprint@dblp#1{\href {http://dblp.uni-trier.de/rec/bibtex/#1.xml}
  {dblp:#1}}
\def\mn@eprint@#1:#2:#3:#4\@nil{\def\@tempa {#1}\def\@tempb {#2}\def\@tempc
  {#3}\ifx \@tempc \@empty \let \@tempc \@tempb \let \@tempb \@tempa \fi \ifx
  \@tempb \@empty \def\@tempb {arXiv}\fi \@ifundefined
  {mn@eprint@\@tempb}{\@tempb:\@tempc}{\expandafter \expandafter \csname
  mn@eprint@\@tempb\endcsname \expandafter{\@tempc}}}

\bibitem[\protect\citeauthoryear{{Baldry}, {Glazebrook}, {Brinkmann},
  {Ivezi{\'c}}, {Lupton}, {Nichol}  \& {Szalay}}{{Baldry}
  et~al.}{2004}]{Baldry2004}
{Baldry} I.~K.,  {Glazebrook} K.,  {Brinkmann} J.,  {Ivezi{\'c}} {\v{Z}}.,
  {Lupton} R.~H.,  {Nichol} R.~C.,   {Szalay} A.~S.,  2004, \mn@doi [\apj]
  {10.1086/380092}, \href
  {https://ui.adsabs.harvard.edu/abs/2004ApJ...600..681B} {600, 681}

\bibitem[\protect\citeauthoryear{{Balogh}, {Baldry}, {Nichol}, {Miller},
  {Bower}  \& {Glazebrook}}{{Balogh} et~al.}{2004}]{Balogh2004}
{Balogh} M.~L.,  {Baldry} I.~K.,  {Nichol} R.,  {Miller} C.,  {Bower} R.,
  {Glazebrook} K.,  2004, \mn@doi [\apjl] {10.1086/426079}, \href
  {https://ui.adsabs.harvard.edu/abs/2004ApJ...615L.101B} {615, L101}

\bibitem[\protect\citeauthoryear{{Bekki}}{{Bekki}}{2009}]{Bekki2009}
{Bekki} K.,  2009, \mn@doi [\mnras] {10.1111/j.1365-2966.2009.15431.x}, \href
  {https://ui.adsabs.harvard.edu/abs/2009MNRAS.399.2221B} {399, 2221}

\bibitem[\protect\citeauthoryear{{Bell} et~al.,}{{Bell}
  et~al.}{2004}]{Bell2004}
{Bell} E.~F.,  et~al., 2004, \mn@doi [\apjl] {10.1086/381388}, \href
  {https://ui.adsabs.harvard.edu/abs/2004ApJ...600L..11B} {600, L11}

\bibitem[\protect\citeauthoryear{{Bell} et~al.,}{{Bell}
  et~al.}{2012}]{Bell2012}
{Bell} E.~F.,  et~al., 2012, \mn@doi [\apj] {10.1088/0004-637X/753/2/167},
  \href {https://ui.adsabs.harvard.edu/abs/2012ApJ...753..167B} {753, 167}

\bibitem[\protect\citeauthoryear{Benson}{Benson}{2012}]{Benson2012}
Benson A.~J.,  2012, \mn@doi [New Astronomy]
  {https://doi.org/10.1016/j.newast.2011.07.004}, 17, 175

\bibitem[\protect\citeauthoryear{{Blanton} et~al.,}{{Blanton}
  et~al.}{2003}]{Blanton2003}
{Blanton} M.~R.,  et~al., 2003, \mn@doi [\apj] {10.1086/375528}, \href
  {https://ui.adsabs.harvard.edu/abs/2003ApJ...594..186B} {594, 186}

\bibitem[\protect\citeauthoryear{{Bower}, {Benson}, {Malbon}, {Helly}, {Frenk},
  {Baugh}, {Cole}  \& {Lacey}}{{Bower} et~al.}{2006}]{Bower2006}
{Bower} R.~G.,  {Benson} A.~J.,  {Malbon} R.,  {Helly} J.~C.,  {Frenk} C.~S.,
  {Baugh} C.~M.,  {Cole} S.,   {Lacey} C.~G.,  2006, \mn@doi [\mnras]
  {10.1111/j.1365-2966.2006.10519.x}, \href
  {https://ui.adsabs.harvard.edu/abs/2006MNRAS.370..645B} {370, 645}

\bibitem[\protect\citeauthoryear{{Boylan-Kolchin}, Springel, White, Jenkins  \&
  Lemson}{{Boylan-Kolchin} et~al.}{2009}]{BoylanKolchin2009}
{Boylan-Kolchin} M.,  Springel V.,  White S. D.~M.,  Jenkins A.,   Lemson G.,
  2009, \mn@doi [\mnras] {10.1111/j.1365-2966.2009.15191.x}, 398, 1150

\bibitem[\protect\citeauthoryear{{Bradford}, {Geha}, {Greene}, {Reines}  \&
  {Dickey}}{{Bradford} et~al.}{2018}]{Bradford2018}
{Bradford} J.~D.,  {Geha} M.~C.,  {Greene} J.~E.,  {Reines} A.~E.,   {Dickey}
  C.~M.,  2018, \mn@doi [\apj] {10.3847/1538-4357/aac88d}, \href
  {https://ui.adsabs.harvard.edu/abs/2018ApJ...861...50B} {861, 50}

\bibitem[\protect\citeauthoryear{{Cattaneo} et~al.,}{{Cattaneo}
  et~al.}{2009}]{Cattaneo2009}
{Cattaneo} A.,  et~al., 2009, \mn@doi [\nat] {10.1038/nature08135}, \href
  {https://ui.adsabs.harvard.edu/abs/2009Natur.460..213C} {460, 213}

\bibitem[\protect\citeauthoryear{{Chabrier}}{{Chabrier}}{2003}]{Chabrier2003}
{Chabrier} G.,  2003, \mn@doi [\pasp] {10.1086/376392}, \href
  {https://ui.adsabs.harvard.edu/abs/2003PASP..115..763C} {115, 763}

\bibitem[\protect\citeauthoryear{{Chadayammuri}, {Tremmel}, {Nagai}, {Babul}
  \& {Quinn}}{{Chadayammuri} et~al.}{2020}]{Chadayammuri2020}
{Chadayammuri} U.,  {Tremmel} M.,  {Nagai} D.,  {Babul} A.,   {Quinn} T.,
  2020, arXiv e-prints, \href
  {https://ui.adsabs.harvard.edu/abs/2020arXiv200106532C} {p. arXiv:2001.06532}

\bibitem[\protect\citeauthoryear{{Cheung} et~al.,}{{Cheung}
  et~al.}{2012}]{Cheung2012}
{Cheung} E.,  et~al., 2012, \mn@doi [\apj] {10.1088/0004-637X/760/2/131}, \href
  {https://ui.adsabs.harvard.edu/abs/2012ApJ...760..131C} {760, 131}

\bibitem[\protect\citeauthoryear{{Cimatti} et~al.,}{{Cimatti}
  et~al.}{2013}]{Cimatti2013}
{Cimatti} A.,  et~al., 2013, \mn@doi [\apjl] {10.1088/2041-8205/779/1/L13},
  \href {https://ui.adsabs.harvard.edu/abs/2013ApJ...779L..13C} {779, L13}

\bibitem[\protect\citeauthoryear{{Conselice}}{{Conselice}}{2014}]{Conselice2014}
{Conselice} C.~J.,  2014, \mn@doi [\araa]
  {10.1146/annurev-astro-081913-040037}, \href
  {https://ui.adsabs.harvard.edu/abs/2014ARA&A..52..291C} {52, 291}

\bibitem[\protect\citeauthoryear{{Crenshaw}, {Schmitt}, {Kraemer}, {Mushotzky}
  \& {Dunn}}{{Crenshaw} et~al.}{2010}]{Crenshaw2010}
{Crenshaw} D.~M.,  {Schmitt} H.~R.,  {Kraemer} S.~B.,  {Mushotzky} R.~F.,
  {Dunn} J.~P.,  2010, \mn@doi [\apj] {10.1088/0004-637X/708/1/419}, \href
  {https://ui.adsabs.harvard.edu/abs/2010ApJ...708..419C} {708, 419}

\bibitem[\protect\citeauthoryear{{Croton} et~al.,}{{Croton}
  et~al.}{2006}]{Croton2006}
{Croton} D.~J.,  et~al., 2006, \mn@doi [\mnras]
  {10.1111/j.1365-2966.2005.09675.x}, \href
  {https://ui.adsabs.harvard.edu/abs/2006MNRAS.365...11C} {365, 11}

\bibitem[\protect\citeauthoryear{{Dav{\'e}}, {Angl{\'e}s-Alc{\'a}zar},
  {Narayanan}, {Li}, {Rafieferantsoa}  \& {Appleby}}{{Dav{\'e}}
  et~al.}{2019}]{Dave2019}
{Dav{\'e}} R.,  {Angl{\'e}s-Alc{\'a}zar} D.,  {Narayanan} D.,  {Li} Q.,
  {Rafieferantsoa} M.~H.,   {Appleby} S.,  2019, \mn@doi [\mnras]
  {10.1093/mnras/stz937}, \href
  {https://ui.adsabs.harvard.edu/abs/2019MNRAS.486.2827D} {486, 2827}

\bibitem[\protect\citeauthoryear{{Dekel} \& {Birnboim}}{{Dekel} \&
  {Birnboim}}{2008}]{Dekel2008}
{Dekel} A.,  {Birnboim} Y.,  2008, \mn@doi [\mnras]
  {10.1111/j.1365-2966.2007.12569.x}, \href
  {https://ui.adsabs.harvard.edu/abs/2008MNRAS.383..119D} {383, 119}

\bibitem[\protect\citeauthoryear{{Di Matteo}, {Springel}  \& {Hernquist}}{{Di
  Matteo} et~al.}{2005}]{DiMatteo2005}
{Di Matteo} T.,  {Springel} V.,   {Hernquist} L.,  2005, \mn@doi [\nat]
  {10.1038/nature03335}, \href
  {https://ui.adsabs.harvard.edu/abs/2005Natur.433..604D} {433, 604}

\bibitem[\protect\citeauthoryear{{Di Matteo}, {Combes}, {Melchior}  \&
  {Semelin}}{{Di Matteo} et~al.}{2007}]{DiMatteo2007}
{Di Matteo} P.,  {Combes} F.,  {Melchior} A.-L.,   {Semelin} B.,  2007, \mn@doi
  [A\&A] {10.1051/0004-6361:20066959}, 468, 61

\bibitem[\protect\citeauthoryear{{Drory}, {Bender}, {Feulner}, {Hopp},
  {Maraston}, {Snigula}  \& {Hill}}{{Drory} et~al.}{2004}]{Drory2004}
{Drory} N.,  {Bender} R.,  {Feulner} G.,  {Hopp} U.,  {Maraston} C.,  {Snigula}
  J.,   {Hill} G.~J.,  2004, \mn@doi [\apj] {10.1086/420781}, \href
  {https://ui.adsabs.harvard.edu/abs/2004ApJ...608..742D} {608, 742}

\bibitem[\protect\citeauthoryear{Ellison, Patton, Simard  \&
  McConnachie}{Ellison et~al.}{2008}]{Ellison2008}
Ellison S.~L.,  Patton D.~R.,  Simard L.,   McConnachie A.~W.,  2008, \mn@doi
  [AJ] {10.1088/0004-6256/135/5/1877}, 135, 1877

\bibitem[\protect\citeauthoryear{{Ellison}, {Mendel}, {Patton}  \&
  {Scudder}}{{Ellison} et~al.}{2013}]{Ellison2013}
{Ellison} S.~L.,  {Mendel} J.~T.,  {Patton} D.~R.,   {Scudder} J.~M.,  2013,
  \mn@doi [\mnras] {10.1093/mnras/stt1562}, \href
  {https://ui.adsabs.harvard.edu/abs/2013MNRAS.435.3627E} {435, 3627}

\bibitem[\protect\citeauthoryear{{Ellison}, {Viswanathan}, {Patton},
  {Bottrell}, {McConnachie}, {Gwyn}  \& {Cuillandre}}{{Ellison}
  et~al.}{2019}]{Ellison2019}
{Ellison} S.~L.,  {Viswanathan} A.,  {Patton} D.~R.,  {Bottrell} C.,
  {McConnachie} A.~W.,  {Gwyn} S.,   {Cuillandre} J.-C.,  2019, \mn@doi
  [\mnras] {10.1093/mnras/stz1431}, \href
  {https://ui.adsabs.harvard.edu/abs/2019MNRAS.487.2491E} {487, 2491}

\bibitem[\protect\citeauthoryear{{Faber} et~al.,}{{Faber}
  et~al.}{2007}]{Faber2007}
{Faber} S.~M.,  et~al., 2007, \mn@doi [\apj] {10.1086/519294}, \href
  {https://ui.adsabs.harvard.edu/abs/2007ApJ...665..265F} {665, 265}

\bibitem[\protect\citeauthoryear{{Fabian}}{{Fabian}}{2012}]{Fabian2012}
{Fabian} A.~C.,  2012, \mn@doi [\araa] {10.1146/annurev-astro-081811-125521},
  \href {https://ui.adsabs.harvard.edu/abs/2012ARA&A..50..455F} {50, 455}

\bibitem[\protect\citeauthoryear{{Gensior} \& {Kruijssen}}{{Gensior} \&
  {Kruijssen}}{2020}]{Gensior2020b}
{Gensior} J.,  {Kruijssen} J.~M.~D.,  2020, \mn@doi [\mnras]
  {10.1093/mnras/staa3453}, \href
  {https://ui.adsabs.harvard.edu/abs/2020MNRAS.tmp.3256G} {}

\bibitem[\protect\citeauthoryear{{Gensior}, {Kruijssen}  \& {Keller}}{{Gensior}
  et~al.}{2020}]{Gensior2020a}
{Gensior} J.,  {Kruijssen} J.~M.~D.,   {Keller} B.~W.,  2020, \mn@doi [\mnras]
  {10.1093/mnras/staa1184}, \href
  {https://ui.adsabs.harvard.edu/abs/2020MNRAS.495..199G} {495, 199}

\bibitem[\protect\citeauthoryear{{Goto}, {Yamauchi}, {Fujita}, {Okamura},
  {Sekiguchi}, {Smail}, {Bernardi}  \& {Gomez}}{{Goto} et~al.}{2003}]{Goto2003}
{Goto} T.,  {Yamauchi} C.,  {Fujita} Y.,  {Okamura} S.,  {Sekiguchi} M.,
  {Smail} I.,  {Bernardi} M.,   {Gomez} P.~L.,  2003, \mn@doi [\mnras]
  {10.1046/j.1365-2966.2003.07114.x}, \href
  {https://ui.adsabs.harvard.edu/abs/2003MNRAS.346..601G} {346, 601}

\bibitem[\protect\citeauthoryear{{Gunn} \& {Gott}}{{Gunn} \&
  {Gott}}{1972}]{Gunn1972}
{Gunn} J.~E.,  {Gott} J.~Richard I.,  1972, \mn@doi [\apj] {10.1086/151605},
  \href {https://ui.adsabs.harvard.edu/abs/1972ApJ...176....1G} {176, 1}

\bibitem[\protect\citeauthoryear{{Hani}, {Gosain}, {Ellison}, {Patton}  \&
  {Torrey}}{{Hani} et~al.}{2020}]{Hani2020}
{Hani} M.~H.,  {Gosain} H.,  {Ellison} S.~L.,  {Patton} D.~R.,   {Torrey} P.,
  2020, \mn@doi [\mnras] {10.1093/mnras/staa459}, \href
  {https://ui.adsabs.harvard.edu/abs/2020MNRAS.493.3716H} {493, 3716}

\bibitem[\protect\citeauthoryear{{Hopkins}, {Hernquist}, {Cox}  \&
  {Kere{\v{s}}}}{{Hopkins} et~al.}{2008}]{Hopkins2008}
{Hopkins} P.~F.,  {Hernquist} L.,  {Cox} T.~J.,   {Kere{\v{s}}} D.,  2008,
  \mn@doi [\apjs] {10.1086/524362}, \href
  {https://ui.adsabs.harvard.edu/abs/2008ApJS..175..356H} {175, 356}

\bibitem[\protect\citeauthoryear{{Ilbert} et~al.,}{{Ilbert}
  et~al.}{2010}]{Ilbert2010}
{Ilbert} O.,  et~al., 2010, \mn@doi [\apj] {10.1088/0004-637X/709/2/644}, \href
  {https://ui.adsabs.harvard.edu/abs/2010ApJ...709..644I} {709, 644}

\bibitem[\protect\citeauthoryear{{Ilbert} et~al.,}{{Ilbert}
  et~al.}{2013}]{Ilbert2013}
{Ilbert} O.,  et~al., 2013, \mn@doi [\aap] {10.1051/0004-6361/201321100}, \href
  {https://ui.adsabs.harvard.edu/abs/2013A&A...556A..55I} {556, A55}

\bibitem[\protect\citeauthoryear{{Jarvis} et~al.,}{{Jarvis}
  et~al.}{2020}]{Jarvis2020}
{Jarvis} M.~E.,  et~al., 2020, \mn@doi [\mnras] {10.1093/mnras/staa2196}, \href
  {https://ui.adsabs.harvard.edu/abs/2020MNRAS.498.1560J} {498, 1560}

\bibitem[\protect\citeauthoryear{{Jogee} et~al.,}{{Jogee}
  et~al.}{2009}]{Jogee2009}
{Jogee} S.,  et~al., 2009, \mn@doi [\apj] {10.1088/0004-637X/697/2/1971}, \href
  {https://ui.adsabs.harvard.edu/abs/2009ApJ...697.1971J} {697, 1971}

\bibitem[\protect\citeauthoryear{{Kauffmann} et~al.,}{{Kauffmann}
  et~al.}{2003}]{Kauffmann2003}
{Kauffmann} G.,  et~al., 2003, \mn@doi [\mnras]
  {10.1046/j.1365-8711.2003.06292.x}, \href
  {https://ui.adsabs.harvard.edu/abs/2003MNRAS.341...54K} {341, 54}

\bibitem[\protect\citeauthoryear{{Kaviraj}, {Kirkby}, {Silk}  \&
  {Sarzi}}{{Kaviraj} et~al.}{2007}]{Kaviraj2007}
{Kaviraj} S.,  {Kirkby} L.~A.,  {Silk} J.,   {Sarzi} M.,  2007, \mn@doi
  [\mnras] {10.1111/j.1365-2966.2007.12475.x}, \href
  {https://ui.adsabs.harvard.edu/abs/2007MNRAS.382..960K} {382, 960}

\bibitem[\protect\citeauthoryear{{Kennicutt}}{{Kennicutt}}{1998}]{Kennicutt1998}
{Kennicutt} Robert~C. J.,  1998, \mn@doi [\apj] {10.1086/305588}, \href
  {https://ui.adsabs.harvard.edu/abs/1998ApJ...498..541K} {498, 541}

\bibitem[\protect\citeauthoryear{{Khalatyan}, {Cattaneo}, {Schramm},
  {Gottl{\"o}ber}, {Steinmetz}  \& {Wisotzki}}{{Khalatyan}
  et~al.}{2008}]{Khalatyan2008}
{Khalatyan} A.,  {Cattaneo} A.,  {Schramm} M.,  {Gottl{\"o}ber} S.,
  {Steinmetz} M.,   {Wisotzki} L.,  2008, \mn@doi [\mnras]
  {10.1111/j.1365-2966.2008.13093.x}, \href
  {https://ui.adsabs.harvard.edu/abs/2008MNRAS.387...13K} {387, 13}

\bibitem[\protect\citeauthoryear{{Knapen}, {Cisternas}  \&
  {Querejeta}}{{Knapen} et~al.}{2015}]{Knapen2015}
{Knapen} J.~H.,  {Cisternas} M.,   {Querejeta} M.,  2015, \mn@doi [\mnras]
  {10.1093/mnras/stv2135}, \href
  {https://ui.adsabs.harvard.edu/abs/2015MNRAS.454.1742K} {454, 1742}

\bibitem[\protect\citeauthoryear{{Koss} et~al.,}{{Koss}
  et~al.}{2020}]{Koss2020}
{Koss} M.~J.,  et~al., 2020, arXiv e-prints, \href
  {https://ui.adsabs.harvard.edu/abs/2020arXiv201015849K} {p. arXiv:2010.15849}

\bibitem[\protect\citeauthoryear{{Larson}, {Tinsley}  \& {Caldwell}}{{Larson}
  et~al.}{1980}]{Larson1980}
{Larson} R.~B.,  {Tinsley} B.~M.,   {Caldwell} C.~N.,  1980, \mn@doi [\apj]
  {10.1086/157917}, \href
  {https://ui.adsabs.harvard.edu/abs/1980ApJ...237..692L} {237, 692}

\bibitem[\protect\citeauthoryear{{Li}, {Kauffmann}, {Heckman}, {Jing}  \&
  {White}}{{Li} et~al.}{2008}]{Li2008}
{Li} C.,  {Kauffmann} G.,  {Heckman} T.~M.,  {Jing} Y.~P.,   {White} S. D.~M.,
  2008, \mn@doi [\mnras] {10.1111/j.1365-2966.2008.13000.x}, \href
  {https://ui.adsabs.harvard.edu/abs/2008MNRAS.385.1903L} {385, 1903}

\bibitem[\protect\citeauthoryear{{Lin} et~al.,}{{Lin} et~al.}{2007}]{Lin2007}
{Lin} L.,  et~al., 2007, \mn@doi [\apjl] {10.1086/517919}, \href
  {https://ui.adsabs.harvard.edu/abs/2007ApJ...660L..51L} {660, L51}

\bibitem[\protect\citeauthoryear{{Lin} et~al.,}{{Lin} et~al.}{2008}]{Lin2008}
{Lin} L.,  et~al., 2008, \mn@doi [\apj] {10.1086/587928}, \href
  {https://ui.adsabs.harvard.edu/abs/2008ApJ...681..232L} {681, 232}

\bibitem[\protect\citeauthoryear{{L{\'o}pez-Sanjuan}
  et~al.,}{{L{\'o}pez-Sanjuan} et~al.}{2013}]{LopezSanjuan2013}
{L{\'o}pez-Sanjuan} C.,  et~al., 2013, \mn@doi [\aap]
  {10.1051/0004-6361/201220286}, \href
  {https://ui.adsabs.harvard.edu/abs/2013A&A...553A..78L} {553, A78}

\bibitem[\protect\citeauthoryear{{Lotz}, {Jonsson}, {Cox}, {Croton}, {Primack},
  {Somerville}  \& {Stewart}}{{Lotz} et~al.}{2011}]{Lotz2011}
{Lotz} J.~M.,  {Jonsson} P.,  {Cox} T.~J.,  {Croton} D.,  {Primack} J.~R.,
  {Somerville} R.~S.,   {Stewart} K.,  2011, \mn@doi [\apj]
  {10.1088/0004-637X/742/2/103}, \href
  {https://ui.adsabs.harvard.edu/abs/2011ApJ...742..103L} {742, 103}

\bibitem[\protect\citeauthoryear{Lu, Mo, Weinberg  \& Katz}{Lu
  et~al.}{2011}]{Lu2011}
Lu Y.,  Mo H.~J.,  Weinberg M.~D.,   Katz N.,  2011, \mn@doi [\mnras]
  {10.1111/j.1365-2966.2011.19170.x}, 416, 1949

\bibitem[\protect\citeauthoryear{Marinacci et~al.,}{Marinacci
  et~al.}{2018}]{Marinacci2018}
Marinacci F.,  et~al., 2018, \mn@doi [\mnras] {10.1093/mnras/sty2206}, 480,
  5113

\bibitem[\protect\citeauthoryear{{Martig}, {Bournaud}, {Teyssier}  \&
  {Dekel}}{{Martig} et~al.}{2009}]{Martig2009}
{Martig} M.,  {Bournaud} F.,  {Teyssier} R.,   {Dekel} A.,  2009, \mn@doi
  [\apj] {10.1088/0004-637X/707/1/250}, \href
  {https://ui.adsabs.harvard.edu/abs/2009ApJ...707..250M} {707, 250}

\bibitem[\protect\citeauthoryear{{McNamara} \& {Nulsen}}{{McNamara} \&
  {Nulsen}}{2007}]{McNamara2007}
{McNamara} B.~R.,  {Nulsen} P.~E.~J.,  2007, \mn@doi [\araa]
  {10.1146/annurev.astro.45.051806.110625}, \href
  {https://ui.adsabs.harvard.edu/abs/2007ARA&A..45..117M} {45, 117}

\bibitem[\protect\citeauthoryear{{Mihos} \& {Hernquist}}{{Mihos} \&
  {Hernquist}}{1996}]{Mihos1996}
{Mihos} J.~C.,  {Hernquist} L.,  1996, \mn@doi [\apj] {10.1086/177353}, \href
  {https://ui.adsabs.harvard.edu/abs/1996ApJ...464..641M} {464, 641}

\bibitem[\protect\citeauthoryear{{Moore}, {Lake}  \& {Katz}}{{Moore}
  et~al.}{1998}]{Moore1998}
{Moore} B.,  {Lake} G.,   {Katz} N.,  1998, \mn@doi [\apj] {10.1086/305264},
  \href {https://ui.adsabs.harvard.edu/abs/1998ApJ...495..139M} {495, 139}

\bibitem[\protect\citeauthoryear{{Muzzin} et~al.,}{{Muzzin}
  et~al.}{2013}]{Muzzin2013}
{Muzzin} A.,  et~al., 2013, \mn@doi [\apjs] {10.1088/0067-0049/206/1/8}, \href
  {https://ui.adsabs.harvard.edu/abs/2013ApJS..206....8M} {206, 8}

\bibitem[\protect\citeauthoryear{Naiman et~al.,}{Naiman
  et~al.}{2018}]{Naiman2018}
Naiman J.~P.,  et~al., 2018, \mn@doi [\mnras] {10.1093/mnras/sty618}, 477, 1206

\bibitem[\protect\citeauthoryear{{Nelson} et~al.,}{{Nelson}
  et~al.}{2015}]{Nelson2015}
{Nelson} D.,  et~al., 2015, \mn@doi [Astronomy and Computing]
  {10.1016/j.ascom.2015.09.003}, \href
  {https://ui.adsabs.harvard.edu/abs/2015A&C....13...12N} {13, 12}

\bibitem[\protect\citeauthoryear{Nelson et~al.,}{Nelson
  et~al.}{2017}]{Nelson2018}
Nelson D.,  et~al., 2017, \mn@doi [\mnras] {10.1093/mnras/stx3040}, 475, 624

\bibitem[\protect\citeauthoryear{{Nelson} et~al.,}{{Nelson}
  et~al.}{2019}]{Nelson2019}
{Nelson} D.,  et~al., 2019, \mn@doi [Computational Astrophysics and Cosmology]
  {10.1186/s40668-019-0028-x}, \href
  {https://ui.adsabs.harvard.edu/abs/2019ComAC...6....2N} {6, 2}

\bibitem[\protect\citeauthoryear{{Pakmor}, {Springel}, {Bauer}, {Mocz},
  {Munoz}, {Ohlmann}, {Schaal}  \& {Zhu}}{{Pakmor} et~al.}{2016}]{Pakmor2016}
{Pakmor} R.,  {Springel} V.,  {Bauer} A.,  {Mocz} P.,  {Munoz} D.~J.,
  {Ohlmann} S.~T.,  {Schaal} K.,   {Zhu} C.,  2016, \mn@doi [\mnras]
  {10.1093/mnras/stv2380}, \href
  {https://ui.adsabs.harvard.edu/abs/2016MNRAS.455.1134P} {455, 1134}

\bibitem[\protect\citeauthoryear{{Patton}, {Grant}, {Simard}, {Pritchet},
  {Carlberg}  \& {Borne}}{{Patton} et~al.}{2005}]{Patton2005}
{Patton} D.~R.,  {Grant} J.~K.,  {Simard} L.,  {Pritchet} C.~J.,  {Carlberg}
  R.~G.,   {Borne} K.~D.,  2005, \mn@doi [\aj] {10.1086/491672}, \href
  {https://ui.adsabs.harvard.edu/abs/2005AJ....130.2043P} {130, 2043}

\bibitem[\protect\citeauthoryear{{Patton}, {Torrey}, {Ellison}, {Mendel}  \&
  {Scudder}}{{Patton} et~al.}{2013}]{Patton2013}
{Patton} D.~R.,  {Torrey} P.,  {Ellison} S.~L.,  {Mendel} J.~T.,   {Scudder}
  J.~M.,  2013, \mn@doi [\mnras] {10.1093/mnrasl/slt058}, \href
  {https://ui.adsabs.harvard.edu/abs/2013MNRAS.433L..59P} {433, L59}

\bibitem[\protect\citeauthoryear{{Patton}, {Qamar}, {Ellison}, {Bluck},
  {Simard}, {Mendel}, {Moreno}  \& {Torrey}}{{Patton}
  et~al.}{2016}]{Patton2016}
{Patton} D.~R.,  {Qamar} F.~D.,  {Ellison} S.~L.,  {Bluck} A. F.~L.,  {Simard}
  L.,  {Mendel} J.~T.,  {Moreno} J.,   {Torrey} P.,  2016, \mn@doi [\mnras]
  {10.1093/mnras/stw1494}, \href
  {https://ui.adsabs.harvard.edu/abs/2016MNRAS.461.2589P} {461, 2589}

\bibitem[\protect\citeauthoryear{{Patton} et~al.,}{{Patton}
  et~al.}{2020}]{Patton2020}
{Patton} D.~R.,  et~al., 2020, \mn@doi [\mnras] {10.1093/mnras/staa913}, \href
  {https://ui.adsabs.harvard.edu/abs/2020MNRAS.494.4969P} {494, 4969}

\bibitem[\protect\citeauthoryear{{Peng} et~al.,}{{Peng}
  et~al.}{2010}]{Peng2010}
{Peng} Y.-j.,  et~al., 2010, \mn@doi [\apj] {10.1088/0004-637X/721/1/193},
  \href {https://ui.adsabs.harvard.edu/abs/2010ApJ...721..193P} {721, 193}

\bibitem[\protect\citeauthoryear{{Peng}, {Lilly}, {Renzini}  \&
  {Carollo}}{{Peng} et~al.}{2012}]{Peng2012}
{Peng} Y.-j.,  {Lilly} S.~J.,  {Renzini} A.,   {Carollo} M.,  2012, \mn@doi
  [\apj] {10.1088/0004-637X/757/1/4}, \href
  {https://ui.adsabs.harvard.edu/abs/2012ApJ...757....4P} {757, 4}

\bibitem[\protect\citeauthoryear{Pillepich et~al.,}{Pillepich
  et~al.}{2017a}]{Pillepich2018a}
Pillepich A.,  et~al., 2017a, \mn@doi [\mnras] {10.1093/mnras/stx2656}, 473,
  4077

\bibitem[\protect\citeauthoryear{Pillepich et~al.,}{Pillepich
  et~al.}{2017b}]{Pillepich2018b}
Pillepich A.,  et~al., 2017b, \mn@doi [\mnras] {10.1093/mnras/stx3112}, 475,
  648

\bibitem[\protect\citeauthoryear{{Planck Collaboration} et~al.,}{{Planck
  Collaboration} et~al.}{2016}]{PlanckCollaboration2016}
{Planck Collaboration} et~al., 2016, \mn@doi [\aap]
  {10.1051/0004-6361/201525830}, \href
  {https://ui.adsabs.harvard.edu/abs/2016A&A...594A..13P} {594, A13}

\bibitem[\protect\citeauthoryear{{Pontzen}, {Tremmel}, {Roth}, {Peiris},
  {Saintonge}, {Volonteri}, {Quinn}  \& {Governato}}{{Pontzen}
  et~al.}{2017}]{Pontzen2017}
{Pontzen} A.,  {Tremmel} M.,  {Roth} N.,  {Peiris} H.~V.,  {Saintonge} A.,
  {Volonteri} M.,  {Quinn} T.,   {Governato} F.,  2017, \mn@doi [\mnras]
  {10.1093/mnras/stw2627}, \href
  {https://ui.adsabs.harvard.edu/abs/2017MNRAS.465..547P} {465, 547}

\bibitem[\protect\citeauthoryear{{Pozzetti} et~al.,}{{Pozzetti}
  et~al.}{2010}]{Pozzetti2010}
{Pozzetti} L.,  et~al., 2010, \mn@doi [\aap] {10.1051/0004-6361/200913020},
  \href {https://ui.adsabs.harvard.edu/abs/2010A&A...523A..13P} {523, A13}

\bibitem[\protect\citeauthoryear{{Rodighiero} et~al.,}{{Rodighiero}
  et~al.}{2011}]{Rodighiero2011}
{Rodighiero} G.,  et~al., 2011, \mn@doi [\apjl] {10.1088/2041-8205/739/2/L40},
  \href {https://ui.adsabs.harvard.edu/abs/2011ApJ...739L..40R} {739, L40}

\bibitem[\protect\citeauthoryear{{Rodighiero} et~al.,}{{Rodighiero}
  et~al.}{2015}]{Rodighiero2015}
{Rodighiero} G.,  et~al., 2015, \mn@doi [\apjl] {10.1088/2041-8205/800/1/L10},
  \href {https://ui.adsabs.harvard.edu/abs/2015ApJ...800L..10R} {800, L10}

\bibitem[\protect\citeauthoryear{{Rodriguez-Gomez} et~al.,}{{Rodriguez-Gomez}
  et~al.}{2015}]{RodriguezGomez2015}
{Rodriguez-Gomez} V.,  et~al., 2015, \mn@doi [\mnras] {10.1093/mnras/stv264},
  \href {https://ui.adsabs.harvard.edu/abs/2015MNRAS.449...49R} {449, 49}

\bibitem[\protect\citeauthoryear{{Rodr{\'\i}guez Montero}, {Dav{\'e}}, {Wild},
  {Angl{\'e}s-Alc{\'a}zar}  \& {Narayanan}}{{Rodr{\'\i}guez Montero}
  et~al.}{2019}]{RodriguezMontero2019}
{Rodr{\'\i}guez Montero} F.,  {Dav{\'e}} R.,  {Wild} V.,
  {Angl{\'e}s-Alc{\'a}zar} D.,   {Narayanan} D.,  2019, \mn@doi [\mnras]
  {10.1093/mnras/stz2580}, \href
  {https://ui.adsabs.harvard.edu/abs/2019MNRAS.490.2139R} {490, 2139}

\bibitem[\protect\citeauthoryear{{Roth}, {Pontzen}  \& {Peiris}}{{Roth}
  et~al.}{2016}]{Roth2016}
{Roth} N.,  {Pontzen} A.,   {Peiris} H.~V.,  2016, \mn@doi [\mnras]
  {10.1093/mnras/stv2375}, \href
  {https://ui.adsabs.harvard.edu/abs/2016MNRAS.455..974R} {455, 974}

\bibitem[\protect\citeauthoryear{{Sanchez}, {Tremmel}, {Werk}, {Pontzen},
  {Christensen}, {Quinn}, {Loebman}  \& {Cruz}}{{Sanchez}
  et~al.}{2020}]{Sanchez2020}
{Sanchez} N.~N.,  {Tremmel} M.,  {Werk} J.~K.,  {Pontzen} A.,  {Christensen}
  C.,  {Quinn} T.,  {Loebman} S.,   {Cruz} A.,  2020, arXiv e-prints, \href
  {https://ui.adsabs.harvard.edu/abs/2020arXiv200905581S} {p. arXiv:2009.05581}

\bibitem[\protect\citeauthoryear{{Schaye} et~al.,}{{Schaye}
  et~al.}{2015}]{Schaye2015}
{Schaye} J.,  et~al., 2015, \mn@doi [\mnras] {10.1093/mnras/stu2058}, \href
  {https://ui.adsabs.harvard.edu/abs/2015MNRAS.446..521S} {446, 521}

\bibitem[\protect\citeauthoryear{Scudder, Ellison, Torrey, Patton  \&
  Mendel}{Scudder et~al.}{2012}]{Scudder2012}
Scudder J.~M.,  Ellison S.~L.,  Torrey P.,  Patton D.~R.,   Mendel J.~T.,
  2012, \mn@doi [\mnras] {10.1111/j.1365-2966.2012.21749.x}, 426, 549

\bibitem[\protect\citeauthoryear{{Shangguan}, {Ho}  \& {Xie}}{{Shangguan}
  et~al.}{2018}]{Shangguan2018}
{Shangguan} J.,  {Ho} L.~C.,   {Xie} Y.,  2018, \mn@doi [\apj]
  {10.3847/1538-4357/aaa9be}, \href
  {https://ui.adsabs.harvard.edu/abs/2018ApJ...854..158S} {854, 158}

\bibitem[\protect\citeauthoryear{{Shangguan}, {Ho}, {Bauer}, {Wang}  \&
  {Treister}}{{Shangguan} et~al.}{2020}]{Shangguan2020}
{Shangguan} J.,  {Ho} L.~C.,  {Bauer} F.~E.,  {Wang} R.,   {Treister} E.,
  2020, \mn@doi [\apj] {10.3847/1538-4357/aba8a1}, \href
  {https://ui.adsabs.harvard.edu/abs/2020ApJ...899..112S} {899, 112}

\bibitem[\protect\citeauthoryear{{Sijacki}, {Springel}, {Di Matteo}  \&
  {Hernquist}}{{Sijacki} et~al.}{2007}]{Sijacki2007}
{Sijacki} D.,  {Springel} V.,  {Di Matteo} T.,   {Hernquist} L.,  2007, \mn@doi
  [\mnras] {10.1111/j.1365-2966.2007.12153.x}, \href
  {https://ui.adsabs.harvard.edu/abs/2007MNRAS.380..877S} {380, 877}

\bibitem[\protect\citeauthoryear{{Somerville} \& {Dav{\'e}}}{{Somerville} \&
  {Dav{\'e}}}{2015}]{Somerville2015}
{Somerville} R.~S.,  {Dav{\'e}} R.,  2015, \mn@doi [\araa]
  {10.1146/annurev-astro-082812-140951}, \href
  {https://ui.adsabs.harvard.edu/abs/2015ARA&A..53...51S} {53, 51}

\bibitem[\protect\citeauthoryear{{Somerville}, {Hopkins}, {Cox}, {Robertson}
  \& {Hernquist}}{{Somerville} et~al.}{2008}]{Somerville2008}
{Somerville} R.~S.,  {Hopkins} P.~F.,  {Cox} T.~J.,  {Robertson} B.~E.,
  {Hernquist} L.,  2008, \mn@doi [\mnras] {10.1111/j.1365-2966.2008.13805.x},
  \href {https://ui.adsabs.harvard.edu/abs/2008MNRAS.391..481S} {391, 481}

\bibitem[\protect\citeauthoryear{{Springel}}{{Springel}}{2010}]{Springel2010}
{Springel} V.,  2010, \mn@doi [\mnras] {10.1111/j.1365-2966.2009.15715.x},
  \href {https://ui.adsabs.harvard.edu/abs/2010MNRAS.401..791S} {401, 791}

\bibitem[\protect\citeauthoryear{{Springel} \& {Hernquist}}{{Springel} \&
  {Hernquist}}{2003}]{Springel2003}
{Springel} V.,  {Hernquist} L.,  2003, \mn@doi [\mnras]
  {10.1046/j.1365-8711.2003.06206.x}, \href
  {https://ui.adsabs.harvard.edu/abs/2003MNRAS.339..289S} {339, 289}

\bibitem[\protect\citeauthoryear{{Springel}, {Di Matteo}  \&
  {Hernquist}}{{Springel} et~al.}{2005a}]{Springel2005b}
{Springel} V.,  {Di Matteo} T.,   {Hernquist} L.,  2005a, \mn@doi [\mnras]
  {10.1111/j.1365-2966.2005.09238.x}, \href
  {https://ui.adsabs.harvard.edu/abs/2005MNRAS.361..776S} {361, 776}

\bibitem[\protect\citeauthoryear{{Springel} et~al.,}{{Springel}
  et~al.}{2005b}]{Springel2005}
{Springel} V.,  et~al., 2005b, \mn@doi [\nat] {10.1038/nature03597}, \href
  {https://ui.adsabs.harvard.edu/abs/2005Natur.435..629S} {435, 629}

\bibitem[\protect\citeauthoryear{{Springel}, {Di Matteo}  \&
  {Hernquist}}{{Springel} et~al.}{2005c}]{Springel2005a}
{Springel} V.,  {Di Matteo} T.,   {Hernquist} L.,  2005c, \mn@doi [\apjl]
  {10.1086/428772}, \href
  {https://ui.adsabs.harvard.edu/abs/2005ApJ...620L..79S} {620, L79}

\bibitem[\protect\citeauthoryear{Springel et~al.,}{Springel
  et~al.}{2017}]{Springel2018}
Springel V.,  et~al., 2017, \mn@doi [\mnras] {10.1093/mnras/stx3304}, 475, 676

\bibitem[\protect\citeauthoryear{{Strateva} et~al.,}{{Strateva}
  et~al.}{2001}]{Strateva2001}
{Strateva} I.,  et~al., 2001, \mn@doi [\aj] {10.1086/323301}, \href
  {https://ui.adsabs.harvard.edu/abs/2001AJ....122.1861S} {122, 1861}

\bibitem[\protect\citeauthoryear{{Tacchella} et~al.,}{{Tacchella}
  et~al.}{2015}]{Tacchella2015}
{Tacchella} S.,  et~al., 2015, \mn@doi [Science] {10.1126/science.1261094},
  \href {https://ui.adsabs.harvard.edu/abs/2015Sci...348..314T} {348, 314}

\bibitem[\protect\citeauthoryear{{Terrazas} et~al.,}{{Terrazas}
  et~al.}{2020}]{Terrazas2020}
{Terrazas} B.~A.,  et~al., 2020, \mn@doi [\mnras] {10.1093/mnras/staa374},
  \href {https://ui.adsabs.harvard.edu/abs/2020MNRAS.493.1888T} {493, 1888}

\bibitem[\protect\citeauthoryear{{Thorp}, {Ellison}, {Simard}, {S{\'a}nchez}
  \& {Antonio}}{{Thorp} et~al.}{2019}]{Thorp2019}
{Thorp} M.~D.,  {Ellison} S.~L.,  {Simard} L.,  {S{\'a}nchez} S.~F.,
  {Antonio} B.,  2019, \mn@doi [\mnras] {10.1093/mnrasl/sly185}, \href
  {https://ui.adsabs.harvard.edu/abs/2019MNRAS.482L..55T} {482, L55}

\bibitem[\protect\citeauthoryear{Villar-Mart\'in, Humphrey, Delgado, Colina  \&
  Arribas}{Villar-Mart\'in et~al.}{2011}]{VillarMartin2011}
Villar-Mart\'in M.,  Humphrey A.,  Delgado R.~G.,  Colina L.,   Arribas S.,
  2011, \mn@doi [\mnras] {10.1111/j.1365-2966.2011.19622.x}, 418, 2032

\bibitem[\protect\citeauthoryear{Vogelsberger, Genel, Sijacki, Torrey, Springel
   \& Hernquist}{Vogelsberger et~al.}{2013}]{Vogelsberger2013}
Vogelsberger M.,  Genel S.,  Sijacki D.,  Torrey P.,  Springel V.,   Hernquist
  L.,  2013, \mn@doi [\mnras] {10.1093/mnras/stt1789}, 436, 3031

\bibitem[\protect\citeauthoryear{{Vogelsberger} et~al.,}{{Vogelsberger}
  et~al.}{2014}]{Vogelsberger2014}
{Vogelsberger} M.,  et~al., 2014, \mn@doi [\mnras] {10.1093/mnras/stu1536},
  \href {https://ui.adsabs.harvard.edu/abs/2014MNRAS.444.1518V} {444, 1518}

\bibitem[\protect\citeauthoryear{{Weigel} et~al.,}{{Weigel}
  et~al.}{2017}]{Weigel2017}
{Weigel} A.~K.,  et~al., 2017, \mn@doi [\apj] {10.3847/1538-4357/aa8097}, \href
  {https://ui.adsabs.harvard.edu/abs/2017ApJ...845..145W} {845, 145}

\bibitem[\protect\citeauthoryear{{Weinberger} et~al.,}{{Weinberger}
  et~al.}{2017}]{Weinberger2017}
{Weinberger} R.,  et~al., 2017, \mn@doi [\mnras] {10.1093/mnras/stw2944}, \href
  {https://ui.adsabs.harvard.edu/abs/2017MNRAS.465.3291W} {465, 3291}

\bibitem[\protect\citeauthoryear{{Weinberger} et~al.,}{{Weinberger}
  et~al.}{2018}]{weinberger2018}
{Weinberger} R.,  et~al., 2018, \mn@doi [\mnras] {10.1093/mnras/sty1733}, \href
  {https://ui.adsabs.harvard.edu/abs/2018MNRAS.479.4056W} {479, 4056}

\bibitem[\protect\citeauthoryear{{Whitaker} et~al.,}{{Whitaker}
  et~al.}{2011}]{Whitaker2011}
{Whitaker} K.~E.,  et~al., 2011, \mn@doi [\apj] {10.1088/0004-637X/735/2/86},
  \href {https://ui.adsabs.harvard.edu/abs/2011ApJ...735...86W} {735, 86}

\bibitem[\protect\citeauthoryear{{White} \& {Frenk}}{{White} \&
  {Frenk}}{1991}]{White1991}
{White} S. D.~M.,  {Frenk} C.~S.,  1991, \mn@doi [\apj] {10.1086/170483}, \href
  {https://ui.adsabs.harvard.edu/abs/1991ApJ...379...52W} {379, 52}

\bibitem[\protect\citeauthoryear{{Willmer} et~al.,}{{Willmer}
  et~al.}{2006}]{Willmer2006}
{Willmer} C.~N.~A.,  et~al., 2006, \mn@doi [\apj] {10.1086/505455}, \href
  {https://ui.adsabs.harvard.edu/abs/2006ApJ...647..853W} {647, 853}

\bibitem[\protect\citeauthoryear{{Wilson} et~al.,}{{Wilson}
  et~al.}{2019}]{Wilson2019}
{Wilson} T.~J.,  et~al., 2019, \mn@doi [\apj] {10.3847/1538-4357/ab06ee}, \href
  {https://ui.adsabs.harvard.edu/abs/2019ApJ...874...18W} {874, 18}

\bibitem[\protect\citeauthoryear{{Wong} et~al.,}{{Wong}
  et~al.}{2011}]{Wong2011}
{Wong} K.~C.,  et~al., 2011, \mn@doi [\apj] {10.1088/0004-637X/728/2/119},
  \href {https://ui.adsabs.harvard.edu/abs/2011ApJ...728..119W} {728, 119}

\bibitem[\protect\citeauthoryear{{Woo}, {Dekel}, {Faber}  \& {Koo}}{{Woo}
  et~al.}{2015}]{Woo2015}
{Woo} J.,  {Dekel} A.,  {Faber} S.~M.,   {Koo} D.~C.,  2015, \mn@doi [\mnras]
  {10.1093/mnras/stu2755}, \href
  {https://ui.adsabs.harvard.edu/abs/2015MNRAS.448..237W} {448, 237}

\bibitem[\protect\citeauthoryear{Woo, Bae, Son  \& Karouzos}{Woo
  et~al.}{2016}]{Woo2016}
Woo J.-H.,  Bae H.-J.,  Son D.,   Karouzos M.,  2016, \mn@doi [ApJ]
  {10.3847/0004-637x/817/2/108}, 817, 108

\bibitem[\protect\citeauthoryear{{Woo}, {Carollo}, {Faber}, {Dekel}  \&
  {Tacchella}}{{Woo} et~al.}{2017}]{Woo2017}
{Woo} J.,  {Carollo} C.~M.,  {Faber} S.~M.,  {Dekel} A.,   {Tacchella} S.,
  2017, \mn@doi [\mnras] {10.1093/mnras/stw2403}, \href
  {https://ui.adsabs.harvard.edu/abs/2017MNRAS.464.1077W} {464, 1077}

\bibitem[\protect\citeauthoryear{{Wuyts} et~al.,}{{Wuyts}
  et~al.}{2011}]{Wuyts2011}
{Wuyts} S.,  et~al., 2011, \mn@doi [\apj] {10.1088/0004-637X/742/2/96}, \href
  {https://ui.adsabs.harvard.edu/abs/2011ApJ...742...96W} {742, 96}

\bibitem[\protect\citeauthoryear{{Zinger} et~al.,}{{Zinger}
  et~al.}{2020}]{Zinger2020}
{Zinger} E.,  et~al., 2020, \mn@doi [\mnras] {10.1093/mnras/staa2607}, \href
  {https://ui.adsabs.harvard.edu/abs/2020MNRAS.499..768Z} {499, 768}

\bibitem[\protect\citeauthoryear{{de Ravel} et~al.,}{{de Ravel}
  et~al.}{2009}]{DeRavel2009}
{de Ravel} L.,  et~al., 2009, \mn@doi [\aap] {10.1051/0004-6361/200810569},
  \href {https://ui.adsabs.harvard.edu/abs/2009A&A...498..379D} {498, 379}

\makeatother
\end{thebibliography}

%\begin{thebibliography}{99}
%\end{thebibliography}

%%%%%%%%%%%%%%%%%%%%%%%%%%%%%%%%%%%%%%%%%%%%%%%%%%

%%%%%%%%%%%%%%%%% APPENDICES %%%%%%%%%%%%%%%%%%%%%

%\appendix



%%%%%%%%%%%%%%%%%%%%%%%%%%%%%%%%%%%%%%%%%%%%%%%%%%

% Don't change these lines
\bsp	% typesetting comment
\label{lastpage}
\end{document}